\documentclass[twocolumn]{style}
\usepackage{graphicx, url, yfonts,epsf,psfrag,amsfonts,amsmath,amssymb,latexsym,cite,bm}

\usepackage{multirow}
\DeclareMathOperator{\E}{E}

\DeclareMathOperator{\opt}{opt}

\DeclareMathOperator{\genie}{genie}
\DeclareMathOperator{\greedy}{greedy}

\DeclareMathOperator{\convex}{convex}
\DeclareMathOperator{\summ}{sum}

{\global\def\putFrag#1#2#3#4{
        \begin{figure}[htbp]
           \begin{center}
              #4
              \epsfxsize=#3cm
              \epsfbox{epsfigures/#1.eps}
           \end{center}
           \caption{#2}
           \label{fig:#1}
        \end{figure}
  }
}

\newcommand{\figref}[1]{Fig.~\ref{fig:#1}}

\newtheorem{theorem}{Theorem}
\newtheorem{proposition}[theorem]{Proposition}
\newtheorem{corollary}[theorem]{Corollary}

\newtheorem{lemma}[theorem]{Lemma}

\begin{document}

\title{Multiuser Scheduling in a Markov-modeled Downlink using Randomly Delayed ARQ Feedback}

\author{Sugumar Murugesan, \textit{Member, IEEE}, Philip Schniter, \textit{Senior Member, IEEE}
and Ness B. Shroff, \textit{Fellow, IEEE}
\thanks{This work was
supported by the NSF CAREER grant 237037, the Office of Naval Research grant N00014-07-1-0209,
NSF grants CNS-0721236, CNS-0626703, ARO W911NF-08-1-0238, CNS-1065136 and CNS-1012700.}
\thanks{Murugesan was with the Department of ECE, The Ohio State University and is currently with the Department of
ECEE, Arizona State University, Schniter is with the Department of ECE, The Ohio State
University and Shroff holds a joint appointment in both the Department of ECE, and the Department of CSE at The Ohio State
University.
(E-mail: sugumar.murugesan@asu.edu, schniter@ece.osu.edu, shroff@ece.osu.edu)}}
%
%
% \author{Sugumar Murugesan, Philip Schniter, Ness B. Shroff \thanks{Murugesan and Schniter are with the Department of %Electrical
% and Computer Engineering, and Shroff holds a joint appointment in
% both the Departments of Electrical and Computer Engineering, and Computer Science and Engineering at The Ohio State
% University, Columbus, OH 43210, USA.
% (E-mail: \{murugess, schniter, shroff\}@ece.osu.edu)}}
%
% %\date{Winter 2010}

\maketitle
\begin{abstract}
In this paper, we focus on the downlink of a cellular system, which corresponds to the bulk of the
data transfer in such wireless systems. We address the problem of opportunistic multiuser scheduling
under imperfect channel state information, by exploiting the memory inherent in the channel. In our
setting, the channel between the base station and each user is modeled by a two-state
Markov chain and the scheduled user sends back an ARQ feedback signal that arrives at the scheduler
with a random delay that is \emph{i.i.d} across users and time. The scheduler indirectly estimates the channel via accumulated delayed-ARQ
feedback and uses this information to make scheduling decisions.
We formulate a throughput maximization problem as a partially observable Markov decision process
(POMDP). For the case of two users in the system, we
show that a greedy policy is \emph{sum throughput optimal} for any distribution on the ARQ feedback delay.
For the case of more than two users, we prove that the greedy policy is suboptimal and demonstrate, via
numerical studies, that it has near optimal performance. We show that the greedy policy can be implemented by a simple algorithm that does not
require the statistics of the underlying Markov channel or the ARQ feedback delay, thus making it
robust against errors in system parameter estimation. Establishing an equivalence between the
two-user system and a genie-aided system, we obtain a simple closed form expression for the sum
capacity of the Markov-modeled downlink. We further derive inner and outer bounds on the capacity
region of the Markov-modeled downlink and tighten these bounds for special cases of the system parameters.\\

\end{abstract}

\section{Introduction}
With the ever increasing demand for high data rates, opportunistic multiuser scheduling,
introduced by Knopp and Humblet in \cite{Knopp}, and defined as \textit{allocating the resources to the user experiencing the
most favorable channel conditions}, has gained immense popularity among wireless network designers.
Opportunistic multiuser scheduling essentially exploits the multiuser diversity in the system and
has motivated several researchers (e.g., \cite{Paulraj}-\cite{Berry})
to study the performance gains obtained by opportunistic
scheduling under various scenarios.
While the \emph{i.i.d} flat fading model is used in these works to model time
varying channels (for a general treatment on opportunistic scheduling with minimal assumptions on
the channel, see \cite{Shroff}), it fails to capture the memory in the channel
observed in realistic scenarios. Hence, more recently, opportunistic scheduling has also been investigated by modeling the 
channels by Markov chains (e.g., \cite{Srikant, Bharghavan, Stoica, Shakkottai, Cao, Andrews}).
However, in these works, the channel state information that is
crucial for the success of any opportunistic scheduling scheme is assumed to
be readily available at the scheduler. This is a simplifying assumption that does not hold in
reality, where a non-trivial amount of resource must be spent in gathering the information on the
channel state. Another line of work (e.g., \cite{Zorzi, Johnston}) attempts to exploit the memory in the
Markov-modeled channels to gather this information. Specifically, Automatic Repeat reQuest (ARQ)
feedback, that is traditionally used for error control (e.g., \cite{Lin, Chang, Rao, Cho}) at the
data link layer, is used to estimate the state of the Markov-modeled channels.

These two lines of work can be combined to create a new design paradigm: exploit multiuser diversity in Markov-modeled
channels (e.g., \cite{Srikant, Bharghavan, Stoica, Shakkottai, Cao, Andrews}) and use the already existing
ARQ feedback mechanism to estimate the state of these Markov-modeled channels (e.g., \cite{Zorzi,
Johnston}). Assuming instantaneous ARQ feedback (i.e., it arrives at the end of the slot) and ON-OFF Markov channel model (the Gilbert-Elliott model \cite{Gilbert}), this problem was addressed in independent works \cite{Zhao, Murugesan}. In \cite{Zhao}, the authors
studied opportunistic spectrum access in a cognitive radio setting --- a setup mathematically
equivalent to the instantaneous ARQ based opportunistic scheduling in a Markov-modeled downlink --- and showed that a simple
greedy scheduling policy is optimal. In \cite{Murugesan}, we directly addressed the instantaneous ARQ based
downlink scheduling problem. By identifying a special mathematical structure in the problem, we
derived a closed form expression for the two-user sum capacity of the downlink and obtained
bounds on the system stability region.

In this paper, we model the downlink channels by two state (ON-OFF) Markov chains and study the ARQ based joint channel learning-scheduling problem when the ARQ feedback arrives
at the scheduler with a \textit{random delay} that is \emph{i.i.d} across users and time. The delay in the feedback channel is an important consideration that cannot be overlooked in realistic scenarios. The effect of feedback delay on channel resource allocation has been studied under various settings in the past (e.g., \cite{Viswanathan, Ying, Annapureddy, Sarkar}). While these works assume deterministic delay, we consider random, \emph{i.i.d} feedback delay. An instance when the feedback delay can be \emph{i.i.d} is when the delay is due to channel propagation time of the feedback signal and when the feedback channel environment changes drastically due to high mobility of users. In essence, by modeling the feedback delay to be random, we attempt to capture the effect of the non-idealities of the feedback channel on the joint channel learning-scheduling problem, in a more general framework.

It turns out that, despite the random delay, the ARQ feedback can be used for opportunistic scheduling to
achieve performance gains. A sample of this gain is
illustrated in \figref{motivate_final} for a specific set of system parameters to be defined in the next
section. \figref{motivate_final} plots the sum (over all the downlink users) \textit{rate} of
successful transmission of packets over a
length of $m$ slots under optimal opportunistic scheduling when the scheduler has: (a) randomly delayed channel state
information (CSI) from \textit{all} the downlink users (b) randomly delayed CSI from the scheduled user --- i.e., randomly delayed ARQ feedback, and
(c) no CSI - i.e., random scheduling. We make two observations from the figure: (1) Using delayed ARQ feedback for opportunistic scheduling can achieve performance close to opportunistic scheduling using delayed CSI from all users, and (2) a
$49\%$ gain (when $m=7$) in the sum rate is associated with opportunistic
scheduling using delayed ARQ over random scheduling. These observations motivate our approach: exploit multiuser diversity in Markov-modeled downlink channels
using the already existing (albeit delayed) ARQ feedback mechanisms.
\putFrag{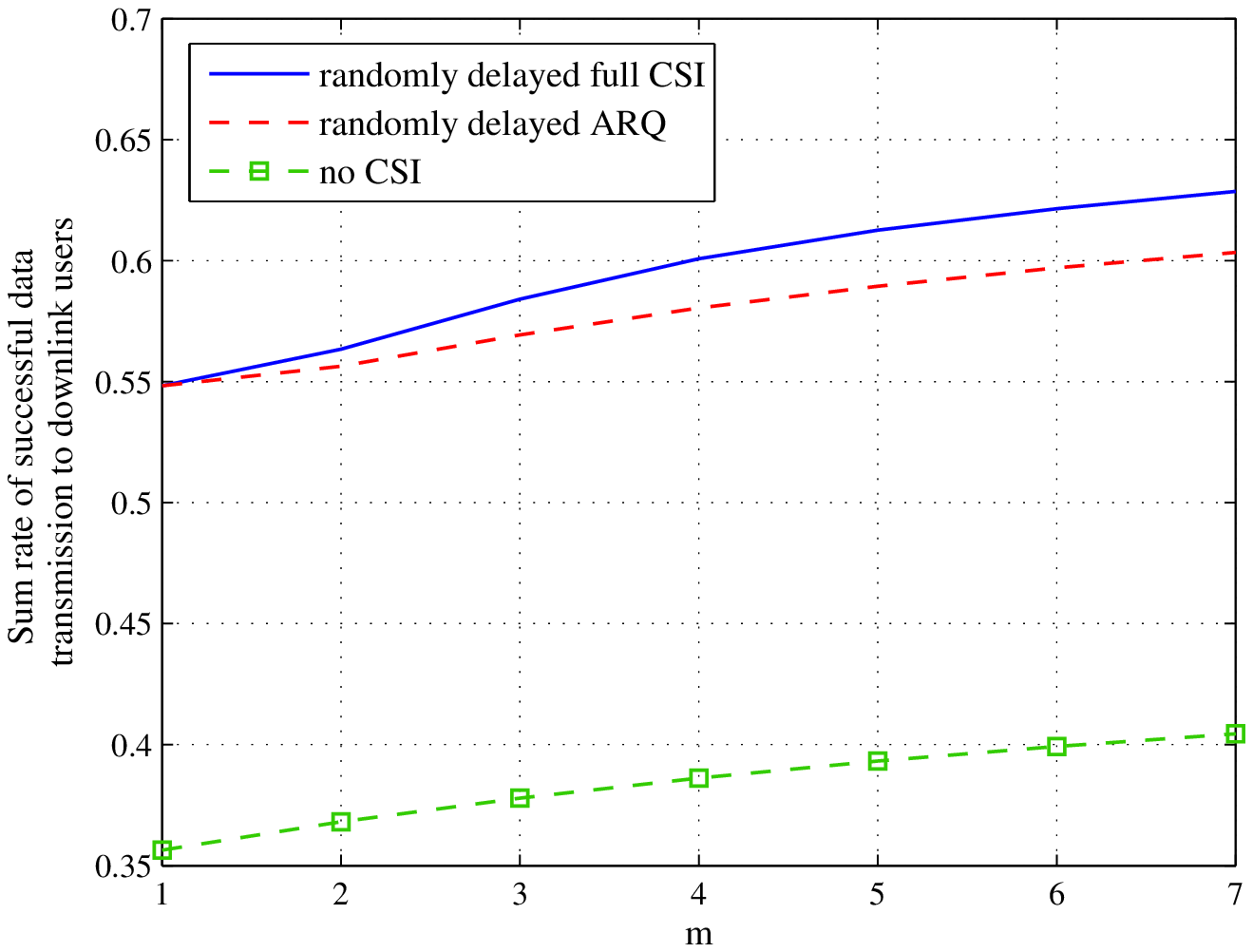}{Illustration of the gains associated with opportunistic
scheduling using randomly delayed ARQ feedback. System parameters used: $p=0.8700$  $r=0.1083$
 $P_D(d=0)=\frac{1}{3}$, $P_D(d=1)=\frac{1}{3}$, $P_D(d=2)=\frac{1}{3}$, $P_D(d>2)=0$,
$\pi_m=[0.3358~0.1851~0.5483]$.}{9}{}

When compared to the instantaneous ARQ case, the randomly delayed ARQ case adds additional layers of complexity to the scheduling problem,
making it different and far more challenging than the former. However, we show that, when there are two users in the system, for any ARQ delay distribution,
the greedy policy that was optimal in the instantaneous ARQ case \cite{Zhao} is also optimal in the delayed ARQ case.
For more than two users, however, using a counterexample, we show that the greedy policy is not, in
general, optimal. Despite the suboptimality, extensive numerical experiments suggest that
the greedy policy has near optimal performance. Encouraged by this insight, we study the structure of the greedy policy and show that it can be implemented via a simple algorithm that is immune to errors in the estimates of the Markov channel parameters and the ARQ delay statistics. We also study the fundamental limits of the Markov-modeled downlink with
randomly delayed ARQ feedback. By establishing an equivalence between the two-user downlink and a
genie-aided system, we derive a simple closed form expression for
the sum capacity of the two-user downlink, while obtaining bounds on
the sum capacity for larger number of users. We further derive inner and outer bounds on the
capacity region of the downlink and tighten these bounds for special cases of the system parameters.

The rest of the paper is organized as follows. The problem setup is described in Section~\ref{sec:sec1}, followed
by a study of the optimality properties of the greedy policy in Section~\ref{sec:optproof}. Section~\ref{sec:greedyperf}
contains a numerical performance analysis of the greedy policy. In Section~\ref{sec:struc}, we discuss the implementation structure of the
greedy policy. We then study the sum capacity and the capacity region of the Markov-modeled downlink in
Section~\ref{sec:limits}, followed by concluding remarks in Section~\ref{sec:conclusion}.

\section{Problem Setup}\label{sec:sec1}

\subsection{Channel Model}

We consider downlink transmissions with $N$ users.
For each user, there is an associated queue at the
base station that accumulates packets intended for that user. We assume that each queue is
infinitely backlogged.
The channel between the base station and each user is modeled
by an \emph{i.i.d} two-state Markov
chain. Each state corresponds to the degree of decodability of the data sent through the channel.
State $1$ (ON) corresponds to full decodability, while state $0$ (OFF) corresponds to zero decodability.
Time is slotted and the channel of each user remains fixed for a slot and
moves into another state in the next slot following the state transition probability of the Markov
chain. The time slots of
all users are synchronized. The two-state Markov channel is characterized by a $2\times 2$ probability transition matrix
\begin{eqnarray}\label{eq:Mmatrix}
P&=&\begin{bmatrix}
p&1-p\\
r&1-r\\
\end{bmatrix},
\end{eqnarray}
where
%\begin{itemize}
\begin{eqnarray}p&:=& \textrm{Prob\{channel is in ON state in the current slot} | \nonumber\\
&&\hspace{26pt}\textrm{channel was in ON state in the previous slot\}}\nonumber\\
r&:=& \textrm{Prob\{channel is in ON state in the current slot} | \nonumber\\
&&\hspace{22pt}\textrm{channel was in OFF state in the previous slot\}}.\nonumber
\end{eqnarray}
%
% \item $r$:= prob(channel is in ON state in the current slot $|$ channel was in OFF state in
% the previous slot)
% \end{itemize}
The states can be interpreted as a quantized representation of the underlying
channel strength, which lies on a continuum. It is known from classic works \cite{Wang0, Zhang0} that the fading channel, with reasonable accuracy, can be modeled by finite state Markov chains and that, in reality, the fading process is observed to be gradual enough that the state transitions/crossovers can be restricted to adjacent states of the Markov model. With the top `half' of the states in these models cumulatively represented by the ON state and the rest by the OFF state in our two-state model, we see that, in realistic scenarios, the crossover from ON to OFF state (respectively, OFF to ON) is less likely to occur than staying in ON state (respectively, OFF state). This is positive correlation, i.e., $p>r$. Motivated by this, we restrict our attention to $p>r$ throughout this work.

\subsection{Scheduling Problem}
The base station (henceforth known as the scheduler) is the central controller that controls the transmission to the users in each slot. In any time
slot, the scheduler does not know the \textit{exact} channel
state of the users and it must schedule the transmission of the head-of-line packet of exactly one
user. Thus, a TDMA styled scheduling is performed here. The power spent in each transmission is
fixed. At the beginning of a time slot, the head-of-line packet of the scheduled user is transmitted. The scheduled user attempts to decode
the received packet and based on the decodability of the packet sends back ACK(bit $1$)/NACK(bit
$0$) feedback
signals to the scheduler at the end of the time slot, over an error-free feedback channel. The
feedback channel is assumed to suffer from a random delay that is \emph{i.i.d} across users and time.
This delayed feedback information, along with the label of the time slot from
which it is acquired, will be used by the scheduler in scheduling decisions. The scheduler aims to
maximize the sum of the rate of successful transmission of packets to all the users in the system. We formally define the
problem below.

\subsection{Formal Problem Definition}\label{subsec:prob defn}
Since the scheduler must make scheduling decisions based only on a partial observation\footnote{In this
case, the set of time-stamped binary delayed feedback on the channels.} of the underlying Markov
chain, the scheduling problem can be represented by a \emph{partially observable Markov decision
process (POMDP)}. See
\cite{Smallwood} for an overview of POMDPs. We now formulate our problem in the language of POMDPs.
The key quantities used throughout this paper are summarized in Appendix~\ref{app:key}.

\textit{Horizon}: The number of consecutive slots over which scheduling is performed is the horizon.
We index the time slots in decreasing order with slot $1$ corresponding to the end of the horizon.
Throughout this paper, the horizon is denoted by $m$, i.e., the scheduling process begins at slot $m$.

\textit{Feedback arriving at slot $t$}: For some slot $t$, $t\le m$, let $n(t)$ be the number of ARQ feedback bits ($\{0,1\}$)
arriving at the end of slot $t$ from the users scheduled in the previous slots. Due to the random nature of the feedback
delay, $n(t)$ can take values in the set $\{0,\ldots,m-t+1\}$. Let $F_t$ represent all the ARQ feedback arriving at the end of
slot $t$. Thus $F_t\in\{0,1\}^{n(t)}$, if $n(t)>0$ and $F_t=\emptyset$, if $n(t)=0$. The ARQ feedback is time-stamped and thus, since the scheduler has a record on which users
were scheduled in the past slots, it can map the feedback bits $F_t$ to the users and slots they
originated from. Let $f_k$ be the feedback that originated during
slot $k$, where $k\le m$. Note that since in each slot one and only one user is scheduled, $f_k$ is
neither empty nor has multiple values, i.e., $f_{k}\in\{0,1\}$ with bit $0$ mapped to NACK
and bit $1$ to ACK feedback.

\textit{Delay of feedback from user $i$ in slot $t$}: Let $D(i,t)$ be the random variable
corresponding to the delay, in number of slots, experienced by the feedback sent by user $i$ in slot $t$. Let
$D(i,t)=0$ correspond to the case when the ARQ feedback originating from user $i$ in
slot $t$ arrives at the scheduler at the end of the same slot $t$. We assume the distribution of $D(i,t)$ to be \emph{i.i.d} across users $i$ and time $t$
throughout this work, and let $P_D(d)$, $d\in\{0,1,\ldots\}$ denote the probability mass function of $D$.

\textit{Belief value of user $i$ in slot $t$ - $\pi_{t}(i)$}: This represents the
probability that the channel of user $i\in\{1\ldots N\}$, in slot $t$, is in the ON state, given all
the past feedback about the channel.
Define $T^u(.)$, for $u\in \{0,1,\ldots\}$, as the $u$-step belief evolution operator given by
$T^u(x)=T(T^{(u-1)}(x))=T^{(u-1)}(T(x))$ with $T(x)=xp+(1-x)r$ and $T^{0}(x)=x$ for $x\in[0,1]$.
Now if, at the end of slot $t+1$, the arriving feedback $F_{t+1}$ contains the ARQ feedback from user
$i$ from slot $k\in\{m, m-1,\ldots,t+1\}$, i.e., $f_k$,
then, if $k$ is the latest slot from which an ARQ feedback from user $i$ has arrived, then
$\pi_t(i)$ is obtained by applying the 1-step belief evolution operator repeatedly over all the
time slots between `now' (slot $t$) and slot $k$, i.e.,
\begin{eqnarray}\label{eq:pievolve}
\pi_{t}(i)=\begin{cases}
T^{k-t}(1)=T^{(k-t-1)}(p),& \mbox{if } f_{k}=1\\
T^{k-t}(0)=T^{(k-t-1)}(r),& \mbox{if } f_{k}=0,
\end{cases}
\end{eqnarray}
where we have used $T^u(x)=T^{u-1}(T(x))$. If $k$ is not the latest slot from which an ARQ feedback from user $i$ has arrived (possible since
the random nature of the feedback delay can result in out-of-turn arrival of ARQ feedback), then due to the first-order Markovian nature of the
channels, this ARQ feedback does not have any new information to affect the belief value, and so
$\pi_t(i)=T(\pi_{t+1}(i))$. Similarly, if $F_{t+1}$ does not contain any feedback from user $i$, then
$\pi_{t}(i)=T(\pi_{t+1}(i))$.

% \textit{Scheduling Policy $\textgoth{A}_{k}$}: A scheduling policy $\textgoth{A}_{k}$ in slot $k$
% is a mapping from the current state of the system to an action as follows:
% \begin{eqnarray}
% \textgoth{A}_{k}:([\pi_{k}],k)\rightarrow a_{k} \quad \forall k\ge 1, \pi_{k}\in [0,1]^N.\nonumber
% \end{eqnarray}
%
\textit{Reward structure}: In any slot $t$, a reward of $1$ is accrued at the
scheduler when the channel of the scheduled user is found to be in the ON state, else $0$ is accrued.

\textit{Scheduling Policy $\textgoth{A}_{k}$}: A scheduling policy $\textgoth{A}_{k}$ in slot $k$ is a
mapping from all the information available at the scheduler in slot $k$ along with the slot index
$k$ to a scheduling decision $a_k$. Formally,
\begin{eqnarray}
\textgoth{A}_{k}:&&\hspace{-10pt}([\pi_m,\pi_{m-1},\ldots,\pi_{k}]^k,\{a_m,a_{m-1},\ldots,a_{k+1}\})\rightarrow
a_{k}\nonumber\\
&&\hspace{-8pt} \forall k \in [1,m], \pi_{k}\in [0,1]^N.
\end{eqnarray}
where $\{a_m,a_{m-1},\ldots,a_{k+1}\}$ are the past scheduling decisions and
$[\pi_m,\pi_{m-1},\ldots,\pi_{k}]^k$ are the belief
values of the channels of all users, corresponding to slots $\{m,m-1,\ldots, k\}$, held by the scheduler
\textit{at the moment} (slot $k$).

\textit{Net expected reward in slot $t$, $V_t$}: With the scheduling policy,
$\{\textgoth{A}_{k}\}_{k=1}^{t}$, fixed,
the net expected reward in slot $t$, i.e., $V_{t}$, is
the sum of the reward expected in the current slot $t$ and the net reward expected in all the future
slots $k<t$. Formally, with $a_k$ denoting the scheduling decision in slot $k$,
\begin{eqnarray}\label{eq:Vm}
\lefteqn{V_t([\pi_m,\pi_{m-1},\ldots,\pi_{t}]^t,\{a_m,a_{m-1},\ldots,a_{t+1}\},\{\textgoth{A}_{k}\}_{k=1}^{t})}\nonumber\\
&=&
R_{t}(\pi_{t},a_t)+\E\big[V_{t-1}([\pi_m,\pi_{m-1},\ldots,\pi_t,\pi_{t-1}]^{t-1},\nonumber\\
&&\hspace{64pt}
\{a_m,a_{m-1},\ldots,a_{t+1},a_{t}\},\{\textgoth{A}_{k}\}_{k=1}^{t-1})\big],\nonumber\\
\end{eqnarray}
where $R_t(\pi_t,a_t)$ is the expected immediate reward and the expectation in the future reward is over the feedback received in slot $t$,
i.e., $F_{t}$, along with the originating slot indices. Note that the belief vector $[\pi_m,\pi_{m-1},\ldots,\pi_{t}]^t$ is up-to-date based on all previous scheduling decisions and the
ARQ feedback received before slot $t$. With the reward structure defined earlier, the expected immediate reward can be written as
\begin{eqnarray}
R_{t}(\pi_{t},a_t)&=&\pi_t(a_t).\nonumber
\end{eqnarray}

\textit{Performance Metric}: For a given scheduling
policy $\{\textgoth{A}_{k}\}_{k=1}^{m}$, the performance metric is given by the sum throughput (sum
rate of successful transmission) over a finite horizon, $m$:
\begin{eqnarray}\label{eq:opt1}
\eta_{\textrm{sum}}(m,\{\textgoth{A}_{k}\}_{k=1}^{m})&=&\frac{V_{m}(\pi_{m},\{\textgoth{A}_{k}\}_{k=1}^{m})}{m},
\end{eqnarray}
where $\pi_{m}$ is the initial belief values of the channels.
%
% Note that, if the reward $\alpha_l$ denotes the
% amount of information decoded when the channel is in state $l$, then the performance metric reflects
% the sum rate of information transferred over the downlink.
%
% \textit{Optimal Scheduling Policy, $\{\textgoth{A}^{*}_{k}\}_{k\ge 1}$}:
% \begin{eqnarray}\label{eq:opt2}
% \{\textgoth{A}^{*}_{k}\}_{k\ge 1}&:=&\arg\max_{\{\textgoth{A}_{k}\}_{k\ge
% 1}}\eta_{\textrm{sum}}(\{\textgoth{A}_{k}\}_{k\ge 1}).
% \end{eqnarray}
%
\section{Greedy Policy - Optimality, Performance Evaluation and the Implementation Structure}\label{sec:sec2}

\subsection{On the Optimality of the Greedy Policy}\label{sec:optproof}

Consider the following policy:
\begin{eqnarray}
\widehat{\textgoth{A}}_{k}: \pi_{k}\rightarrow a_{k}&=&\hspace{-5pt}\arg\max_{i}R_{k}(\pi_{k},a_k=i)\nonumber\\
&=&\hspace{-5pt}\arg\max_{i}\pi_{k}(i) \quad \forall k\ge 1, \pi_{k}\in [0,1]^N.\nonumber\\
\end{eqnarray}
Since the above given policy attempts to maximize the expected immediate reward, without any regard to
the expected future reward, it follows an approach that is fundamentally \textit{greedy} in nature.
We henceforth call $\{\widehat{\textgoth{A}}_{k}\}_{k=1}^{m}$ the greedy policy and let $\hat{a}_k$
denote the scheduling decision in slot $k$ under the greedy policy. We now proceed to
establish the optimality of the greedy policy when $N=2$. We first introduce the following lemma. 
\begin{lemma}\label{le:Tprop}
For any $u,v\in\{0,1,2,\ldots\}$ and any
$x,y\in [0,1]$ with $x\ge y$,
\begin{eqnarray}
%T^u(p)&\ge&T^{u+1}(p)\nonumber\\
T^{u}(p)&\ge&T^{u+1}(x)\nonumber\\
T^{u}(r)&\le&T^{u+1}(x)\nonumber\\
T^{u}(x)&\ge&T^{u}(y)\nonumber\\
T^u(p)&\ge&T^v(r).
%T^u(r)&\le&T^{u+1}(r)\nonumber\\
\end{eqnarray}
\end{lemma}

The results of Lemma~\ref{le:Tprop} can be explained intuitively. Note that $T^u(x)$ is the belief value
of the channel (probability that the channel is in the ON-state) in the current slot given the belief
value, $u$ slots earlier, was $x$. Also note that $T^u(p)$ (similarly $T^u(r)$) gives the belief value in
the current slot given the channel was in the ON state (similarly OFF state) $u+1$ slots earlier. Now,
since the Markov channel is positively correlated ($p>r$), the probability that the channel is in
the ON state in the current slot given it was in the ON state $u+1$ slots earlier ($T^u(p)$) is at least as high as the probability that
the channel is ON in the current slot given it was \textit{ON with probability $x\in[0,1]$}, $u+1$
slots earlier ($T^{(u+1)}(x)$). This explains the first inequality in Lemma~\ref{le:Tprop}.
The second and third inequalities can be explained along similar lines. Regarding the last inequality, consider slots $t,k$ such that $t>k$. Due to
the Markovian nature of the channel, the closer slot $k$ is to $t$, the
stronger is the memory, i.e., the dependency of the channel state in $k$ with that of $t$. Now,
since the channel is positively correlated, if the channel was in the ON state in slot $t$, the closer $k$ is to $t$, the higher is the probability
that the channel is ON in slot $k$. By definition, this
probability is given by $T^u(p)$ with $u=t-k-1$. Thus $T^u(p)$ monotonically decreases with
$u$. Using a similar explanation, $T^u(r)$ monotonically \textit{increases} with $u$. The limiting value of both these functions, as
$u\rightarrow \infty$, is the probability that the channel is ON when no information on the
past channel states is available. This is given by the steady state probability\footnote{We will discuss the steady state probability in
Section~\ref{sec:limits}.}. This explains $T^u(p)\ge T^v(r)$ for any $u,v\in\{0,1,\ldots\}$.
A formal proof of Lemma~\ref{le:Tprop} can be found in Appendix~\ref{app:lemma}.

\begin{proposition}\label{prop:myoopt}
For $N=2$, the sum throughput, $\eta_{\textrm{sum}}(m,\{\textgoth{A}_{k}\}_{k=1}^{m})$, of the system is maximized by the greedy policy
$\{\widehat{\textgoth{A}}_k\}_{k=1}^{m}$ for any ARQ delay distribution.
% \begin{eqnarray}
% \textgoth{A}^{*}_{k}|_{N=2}&=&\widehat{\textgoth{A}}_{k}\quad\forall k\ge 1.\nonumber
% \end{eqnarray}
\end{proposition}

\begin{proof}
Consider a slot $t<m$. Fix a sequence of
scheduling decisions $\vec{a}_{t+1}:=\{a_m,a_{m-1},\ldots,a_{t+1}\}$. 
Recall the definition of $F_{t+1}$, the feedback arriving at the end of slot $t+1$, from Section~\ref{subsec:prob defn}.
% 
% 
% 
% 
% 
% 
% 
% % We now proceed to show that,
% % if the greedy policy is implemented in slot $t$, then the expected immediate reward in slot $t$ is
% % independent of the schedule decisions $\vec{a}_{t+1}$, i.e.,
% % \begin{eqnarray}
% % E_{\pi_{t}|\pi_m,\vec{a}_{t+1}}R_{t}(\pi_t,\hat{a}_t)&=&E_{\pi_{t}|\pi_m}R_{t}(\pi_t,\hat{a}_t).
% % \end{eqnarray}
% % %If the feedback is ACK(NACK) $f_{k}=1(0)$. 
% % When there is no feedback arriving at slot $k$, then
% % $F_k=\emptyset$. If there are multiple feedback arriving at slot $k$, then $F_k$ is a vector. 
% % 
% 
% 
% 
% 
Let $\tau_{t+1}$ denote the originating slots corresponding to feedback $F_{t+1}$, i.e., if the feedback
from users $a_{u}$ and $a_{v}$, for $m\ge u>v\ge t+1$, both arrive at slot $t+1$, then
$F_{t+1}=[f_{u}~f_{v}]$ and $\tau_{t+1}=[u~ v]$. 
Also define $k_1\in\{\emptyset,m,m-1,\ldots,t+1\}$ as the latest slot from which the ARQ feedback of user
$1$ is available at the scheduler by (the beginning of) slot $t$.
% and $l_1=k_1-t-1$ ($l_2=k_2-t-1$) be a measure
% of `freshness' of the latest feedback from user $1$ (user $2$). 
Formally, if at least one ARQ feedback from user 1 has arrived at the scheduler by slot $t$, then 
\begin{eqnarray}
k_1&=&\min_{k\in\{m,m-1,\ldots,t+1\}~\textrm{s.t}~ a_{k}=1,~\textrm{$f_k$~has arrived by slot
$t$}}k.\nonumber\\
%l_1&=&k_1-t-1.
\end{eqnarray}
If no ARQ feedback from user $1$ has arrived by slot $t$, i.e., if $\nexists$ a $k$ such that `$k\in\{m,m-1,\ldots,t+1\}~\textrm{s.t}~ a_{k}=1,~\textrm{$f_k$~has
arrived by slot $t$}$', then $k_1=\emptyset$. Let $l_1=k_1-t-1$, when $k_1\neq \emptyset$, be a measure of `freshness' of the latest
feedback from user $1$. Let $l_1=\emptyset$ when $k_1=\emptyset$. Similarly define $k_2,l_2$ for
user $2$. With these definitions, the proof proceeds in two steps: In step $1$, we show that the greedy decision in slot
$t$, given the ARQ feedback and the scheduling decision from slot $\min(k_1,k_2)$, is independent of the
feedback and scheduling decision corresponding to slot $\max(k_1,k_2)$. 
In step $2$, we show that, if the greedy policy is implemented in slot $t$, then the expected immediate reward in slot $t$ is
independent of the scheduling decisions $\vec{a}_{t+1}$. We then provide induction based arguments to
establish the proposition. 

\textit{Step 1:} Let $\vec{F}_{t+1}:=\{F_m,F_{m-1},\ldots,F_{t+1}\}$ and
$\vec{\tau}_{t+1}:=\{\tau_{m},\tau_{m-1},\ldots,\tau_{t+1}\}$. The greedy decision in slot $t$, conditioned on
the past feedback and scheduling decisions is given by
\begin{eqnarray}
\hat{a}_{t}|_{\vec{F}_{t+1},\vec{\tau}_{t+1},\vec{a}_{t+1},\pi_m}&=&\hat{a}_{t}|_{f_{k_1},f_{k_2},l_1,l_2,\vec{a}_{t+1},\pi_m}.
\end{eqnarray}
The preceding equation comes directly from the first order Markovian property of the underlying
channels.
% ollowing argument: $\hat{a}_t$ is a function of $\pi_t$. $\pi_t(1)$ (likewise,
% $\pi_t(2)$) is independent of $\{_{k\neq k_1},\tau_{k\neq k_1^*}\}$ (likewise, $\{f_{k\neq
% k_2^*},\tau_{k\neq k_2^*}\}$) given $\{f_{k_1},\tau_{k_1^*},\vec{a}_{t+1}\}$
% (likewise, $\{f_{k_2^*},\tau_{k_2^*},\vec{a}_{t+1}\}$) due to the first order Markovian property of the downlink
% channel. 
Consider the case when $k_1<k_2\le m$ ($\Rightarrow l_1<l_2$) or $k_1=k_2= \emptyset$ ($\Rightarrow
l_1=l_2=\emptyset$). The belief values in slot $t$ as a function of feedback $f_{k_1}$ and
$f_{k_2}$ is given below:
\begin{eqnarray}
\lefteqn{(\pi_t(1),\pi_t(2))}\nonumber\\
&=&\hspace{-10pt}\begin{cases}
\hspace{-2pt}(T^{{l_1}}(p),T^{{l_2}}(p)), & \hspace{-7pt} \mbox{if } f_{k_1}=1,f_{k_2}=1 \\
\hspace{-2pt}(T^{{l_1}}(p),T^{{l_2}}(r)), & \hspace{-7pt} \mbox{if } f_{k_1}=1,f_{k_2}=0 \\
\hspace{-2pt}(T^{{l_1}}(p),T^{(m-t)}(\pi_m(2))), & \hspace{-7pt} \mbox{if } f_{k_1}=1,k_2=\emptyset\\
\hspace{-2pt}(T^{{l_1}}(r),T^{{l_2}}(p)), & \hspace{-7pt} \mbox{if } f_{k_1}=0,f_{k_2}=1\\
\hspace{-2pt}(T^{{l_1}}(r),T^{{l_2}}(r)), & \hspace{-7pt} \mbox{if } f_{k_1}=0,f_{k_2}=0\\
\hspace{-2pt}(T^{{l_1}}(r),T^{(m-t)}(\pi_m(2))), & \hspace{-7pt} \mbox{if } f_{k_1}=0,k_2=\emptyset\\
\hspace{-2pt}(T^{(m-t)}(\pi_m(1)),T^{(m-t)}(\pi_m(2))), & \hspace{-7pt} \mbox{if } k_1=\emptyset,k_2=\emptyset\\
\end{cases}\nonumber\\
\end{eqnarray}
Using Lemma~\ref{le:Tprop}, the greedy decision can be written as
\begin{eqnarray}\label{eq:ahat}
\lefteqn{\hat{a}_{t}|_{f_{k_1},f_{k_2},l_1,l_2,\vec{a}_{t+1},\pi_m}}\nonumber\\
&=&\begin{cases}
1, & \mbox{if }f_{k_1}=1\\
2, & \mbox{if }f_{k_1}=0\\
\arg\max_{i\in\{1,2\}}(\pi_m(i)), & \mbox{if } k_1=\emptyset,k_2=\emptyset.
\end{cases}
\end{eqnarray}
Thus the greedy decision is independent of feedback $f_{k_2}$ if $k_1<k_2$.
We now proceed to generalize equation (\ref{eq:ahat}). Let $k^*$ denote the latest slot for which an ARQ feedback is available from
\textit{one of the users} by slot $t$, i.e.,
\begin{eqnarray}\label{eq:kldefn}
k^*&=&\begin{cases}
\min\{k_1,k_2\}, & \mbox{if } k_1\neq\emptyset,k_2\neq\emptyset\\
k_1, & \mbox{if } k_1\neq\emptyset,k_2=\emptyset\\
k_2, & \mbox{if } k_1=\emptyset,k_2\neq\emptyset\\
\emptyset, & \mbox{if } k_1=\emptyset,k_2=\emptyset.
\end{cases}
\end{eqnarray}
Let $l=k^*-t-1$ for $k^*\neq \emptyset$ and $l=\emptyset$ for $k^*=\emptyset$ be a measure of
freshness of the latest ARQ feedback. Thus, using the preceding
discussion, we have
\begin{eqnarray}\label{eq:greedyk}
\lefteqn{\hat{a}_{t}|_{f_{k_1},f_{k_2},l_1,l_2,\vec{a}_{t+1},\pi_m}}\nonumber\\
&=&\hat{a}_{t}|_{f_{k^*},l,\vec{a}_{t+1},\pi_m}\nonumber\\
&=&\begin{cases} a_{k^*}, & \mbox{if } k^*\neq\emptyset,f_{k^*}=1\\
\bar{a}_{k^*}, & \mbox{if } k^*\neq\emptyset,f_{k^*}=0\\
\arg\max_{i\in\{1,2\}}(\pi_m(i)), & \mbox{if } k^*=\emptyset\\
\end{cases}\nonumber\\
\end{eqnarray}
where $\bar{a}_{k^*}$ is the user \textit{not} scheduled in slot $k^*$. This completes step $1$ of the
proof.

% 
% Using the preceding discussion, with
% \begin{eqnarray}
% k&=&\begin{cases}
% \min\{k_1,k_2\}, & \mbox{if } k_1\neq\emptyset,k_2\neq\emptyset\\
% k_1, & \mbox{if } k_1\neq\emptyset,k_2=\emptyset\\
% k_2, & \mbox{if } k_1=\emptyset,k_2\neq\emptyset\\
% \emptyset, & \mbox{if } k_1=\emptyset,k_2=\emptyset,
% \end{cases}
% \end{eqnarray}
% $l=k-t-1$ for $k\neq \emptyset$ and $l=\emptyset$ for $k=\emptyset$, we have
% \begin{eqnarray}
% \hat{a}_{t}|_{f_{k_1},f_{k_2},l_1,l_2,\vec{a}_{t+1},\pi_m}&=&\begin{cases}
% \hat{a}_{t}|_{f_{k},l,\vec{a}_{t+1},\pi_m}, & \mbox{if } k\neq\emptyset\\
% \hat{a}_{t}|_{\pi_m}, & \mbox{if } k=\emptyset.
% \end{cases}\nonumber\\
% \end{eqnarray}
% Note that $k$ is the latest slot for which an ARQ feedback is available from
% \textit{one of the users} by slot $t$.
% 
\textit{Step 2:}  
%We now proceed to show that,
If the greedy policy is implemented in slot $t$, 
%then the expected immediate reward in slot $t$ is
% independent of the scheduling decisions $\vec{a}_{t+1}$, i.e.,
% \begin{eqnarray}
% E_{\pi_{t}|\vec{a}_{t+1},\pi_m}R_{t}(\pi_t,\hat{a}_t)&=&E_{\pi_{t}|\pi_m}R_{t}(\pi_t,\hat{a}_t).
% \end{eqnarray}
the immediate reward expected in slot $t$, conditioned on scheduling decisions $\vec{a}_{t+1}$ and
initial belief $\pi_m$ can be rewritten as
\begin{eqnarray}\label{eq:exprewt}
\lefteqn{\E_{\pi_t|{\vec{a}_{t+1},\pi_m}}R_t(\pi_t,\hat{a}_t)}\nonumber\\
&=&\E_{\pi_t|{l=\emptyset,\vec{a}_{t+1},\pi_m}}(R_t(\pi_t,\hat{a}_t))P(l=\emptyset|{\vec{a}_{t+1},\pi_m})\nonumber\\
&&+\E_{l,l\neq\emptyset|{\vec{a}_{t+1},\pi_m}}\E_{\pi_t|{l,l\neq\emptyset,\vec{a}_{t+1},\pi_m}}(R_t(\pi_t,\hat{a}_t)),
\end{eqnarray}
where $l$ is defined after (\ref{eq:kldefn}). Note that 
\begin{eqnarray}
\E_{\pi_t|{l=\emptyset,\vec{a}_{t+1},\pi_m}}(R_t(\pi_t,\hat{a}_t))&=&\max_{i}T^{(m-t)}(\pi_m(i))
\end{eqnarray}
since, with $l=\emptyset$, i.e., no past feedback at the scheduler, the belief values at slot $t$ is independent of the past
scheduling decisions and is simply given by $\pi_t=T^{(m-t)}(\pi_m)$. Now rewriting the second part of
(\ref{eq:exprewt}),
\begin{eqnarray}\label{eq:EE}
\lefteqn{\E_{l,l\neq\emptyset|{\vec{a}_{t+1},\pi_m}}\E_{\pi_t|{l,l\neq\emptyset,\vec{a}_{t+1},\pi_m}}(R_t(\pi_t,\hat{a}_t))}\nonumber\\
&=&\E_{l,l\neq\emptyset|{\vec{a}_{t+1},\pi_m}}
\E_{\pi_{l+t+1}|{l,l\neq\emptyset,\vec{a}_{t+1},\pi_m}}\nonumber\\
&&\E_{\pi_t|{\pi_{l+t+1},l,l\neq\emptyset,\vec{a}_{t+1},\pi_m}}(R_t(\pi_t,\hat{a}_t)).
\end{eqnarray}
Consider $\E_{\pi_t|{\pi_{l+t+1},l,l\neq\emptyset,\vec{a}_{t+1},\pi_m}}(R_t(\pi_t,\hat{a}_t))$. From
the first step of the proof, the greedy decision in slot $t$ can be made solely based on the
latest feedback, i.e., $f_{k^*=l+t+1}$. This was recorded in (\ref{eq:greedyk}). Thus, if the
feedback $f_{k^*}$ is an ACK (occurs with probability $\pi_{l+t+1}(a_{l+t+1})$) reschedule the user $a_{l+t+1}$ in slot
$t$. Conditioned on $f_{k^*}=1$, the belief value $\pi_{t}(a_{l+t+1})$ and hence the expected
immediate reward in slot $t$ is given by $T^l(p)$. If the feedback is a NACK, schedule the other
user denoted by $\bar{a}_{l+t+1}$. Conditioned on $f_{k^*}=0$, the belief value $\pi_{t}(\bar{a}_{l+t+1})$ and hence the 
expected immediate reward in slot $t$ is given by
$T^{(l+1)}(\pi_{l+t+1}(\bar{a}_{l+t+1}))=\pi_{l+t+1}(\bar{a}_{l+t+1})T^{l}(p)+(1-\pi_{l+t+1}(\bar{a}_{l+t+1}))T^{l}(r)$.
Averaging over $f_{k^*=l+t+1}$, we have
\begin{eqnarray}
\lefteqn{\E_{\pi_t|{\pi_{l+t+1},l,l\neq\emptyset,\vec{a}_{t+1},\pi_m}}(R_t(\pi_t,\hat{a}_t))}\nonumber\\
&=&\pi_{l+t+1}(a_{l+t+1})T^{l}(p)+(1-\pi_{l+t+1}(a_{l+t+1}))\times\nonumber\\
&&\Big(\pi_{l+t+1}(\bar{a}_{l+t+1})T^{l}(p)+(1-\pi_{l+t+1}(\bar{a}_{l+t+1}))T^{l}(r)\Big)\nonumber\\
&=&P\big(\{S_{l+t+1}(1)=1 \cup S_{l+t+1}(2)=1\}|\nonumber\\
&&\hspace{96pt}{\pi_{l+t+1},l,l\neq\emptyset,\vec{a}_{t+1},\pi_m}\big)T^{l}(p)\nonumber\\
&&\hspace{-10pt}+P\big(\{S_{l+t+1}(1)=0 \cap S_{l+t+1}(2)=0\}|\nonumber\\
&&\hspace{96pt}{\pi_{l+t+1},l,l\neq\emptyset,\vec{a}_{t+1},\pi_m}\big)T^{l}(r)\nonumber\\
\end{eqnarray}
where $S_{k}(i)$ is the $1/0$ state of the channel of user $i$ in slot $k$. From (\ref{eq:EE}),
\begin{eqnarray}\label{eq:rewfinal}
\lefteqn{\E_{l,l\neq\emptyset|{\vec{a}_{t+1},\pi_m}}\E_{\pi_t|{l,l\neq\emptyset,\vec{a}_{t+1},\pi_m}}(R_t(\pi_t,\hat{a}_t))}\nonumber\\
&=&\E_{l,l\neq\emptyset|{\vec{a}_{t+1},\pi_m}}\E_{\pi_{l+t+1}|{l,l\neq\emptyset,\vec{a}_{t+1},\pi_m}}\nonumber\\
&&\hspace{-8pt}\Big(P\big(\{S_{l+t+1}(1)=1 \cup S_{l+t+1}(2)=1\}|\nonumber\\
&&\hspace{88pt}{\pi_{l+t+1},l,l\neq\emptyset,\vec{a}_{t+1},\pi_m}\big)T^{l}(p)\nonumber\\
&&\hspace{-10pt}+P\big(\{S_{l+t+1}(1)=0 \cap S_{l+t+1}(2)=0\}|\nonumber\\
&&\hspace{88pt}{\pi_{l+t+1},l,l\neq\emptyset,\vec{a}_{t+1},\pi_m}\big)T^{l}(r)\Big)\nonumber\\
&=&\E_{l,l\neq\emptyset|{\vec{a}_{t+1},\pi_m}}\Big(P\big(\{S_{l+t+1}(1)=1 \cup
S_{l+t+1}(2)=1\}|\nonumber\\
&&\hspace{121pt}{l,l\neq\emptyset,\vec{a}_{t+1},\pi_m}\big)T^{l}(p)\nonumber\\
&&\hspace{53pt}+P\big(\{S_{l+t+1}(1)=0 \cap S_{l+t+1}(2)=0\}|\nonumber\\
&&\hspace{121pt}{l,l\neq\emptyset,\vec{a}_{t+1},\pi_m}\big)T^{l}(r)\Big)\nonumber\\
&=&\E_{l,l\neq\emptyset|{\vec{a}_{t+1},\pi_m}}\nonumber\\
&&\hspace{-8pt}\Big(P\big(\{S_{l+t+1}(1)=1 \cup S_{l+t+1}(2)=1\}|{\pi_m}\big)T^{l}(p)\nonumber\\
&&\hspace{-8pt}+P\big(\{S_{l+t+1}(1)=0 \cap S_{l+t+1}(2)=0\}|{\pi_m}\big)T^{l}(r)\Big)\nonumber\\
\end{eqnarray}
We have used the following argument in the last equality: the event $(\{S_{l+t+1}(1)=1 \cup S_{l+t+1}(2)=1\})$ is
controlled by the underlying Markov dynamics and is independent of the scheduling decisions
$\vec{a}_{t+1}$. Likewise, this event is independent of the value of $l$ since we have assumed that
the feedback channel and the forward channel are independent. 

Recall $D(i,k)$ is the random variable
indicating the delay incurred by the ARQ feedback sent by user $i$ in slot $k$. Let $L$ be the
random variable corresponding to the quantity $l$, the degree of freshness of the latest ARQ
feedback, and $P_L(.)$ be the probability mass function of $L$. Therefore, for $0\le l\le m-t-1$,
\begin{eqnarray}\label{eq:ARQDel}
\lefteqn{P_L(l|\vec{a}_{t+1},\pi_m)}\nonumber\\
&=&P\big(\{D(a_{l+t+1},l+t+1)\le l,D(a_{l+t},l+t)>(l-1),\nonumber\\
& &\hspace{17pt}D(a_{l+t-1},l+t-1)>(l-2)),\ldots,\nonumber\\
&&\hspace{95pt}D(a_{t+1},t+1)>0\}|\vec{a}_{t+1},\pi_m\big)\nonumber\\
&=&P\big(\{D(a_{l+t+1},l+t+1)\le l,D(a_{l+t},l+t)>(l-1),\nonumber\\
& &\hspace{17pt}D(a_{l+t-1},l+t-1)>(l-2)),\ldots,\nonumber\\
&&\hspace{112pt}D(a_{t+1},t+1)>0\}|\vec{a}_{t+1}\big)\nonumber\\
&=&P(D(1,l+t+1)\le l)\prod_{k=t+l}^{t+1}P(D(1,k)> k-t-1)\nonumber\\
\end{eqnarray}
where we have used the independence between the forward and the feedback channel to remove
the condition on $\pi_m$ in the second equality. The last equality comes from the assumption that the ARQ delay is \emph{i.i.d}
across users and time\footnote{Note: here we do not require the ARQ delay to be identically distributed
across time.}. Similarly
\begin{eqnarray}
P_L(l=\emptyset|\vec{a}_{t+1},\pi_m)&=&\prod_{k=m}^{t+1}P(D(a_k,k)>k-t-1)\nonumber\\
\end{eqnarray}
Applying the preceding equations in (\ref{eq:exprewt}), we
have
\begin{eqnarray}
\lefteqn{\E_{\pi_t|{\vec{a}_{t+1},\pi_m}}R_t(\pi_t,\hat{a}_t)}\nonumber\\
&=&\hspace{-7pt}\prod_{k=m}^{t+1}P(D(a_k,k)>k-t-1)\max_{i}T^{(m-t)}(\pi_m(i))\nonumber\\
&&\hspace{-14pt}+\hspace{-7pt}\sum_{l=0}^{m-t-1}\hspace{-2pt}P(D(1,l+t+1)\le
l)\hspace{-7pt}\prod_{k=t+l}^{t+1}\hspace{-5pt}P(D(1,k)>k-t-1)\nonumber\\
&&\hspace{0pt}\Big(P\big(\{S_{l+t+1}(1)=1 \cup S_{l+t+1}(2)=1\}|{\pi_m}\big)T^{l}(p)\nonumber\\
&&\hspace{4pt}+P\big(\{S_{l+t+1}(1)=0 \cap S_{l+t+1}(2)=0\}|{\pi_m}\big)T^{l}(r)\Big)\nonumber\\
\end{eqnarray}
The expected reward in slot $t$ is thus independent of the sequence of actions $\{a_{m},a_{m-1}\ldots
a_{t+1}\}$ if the greedy policy is implemented in slot $t$. By extension, the total reward expected from slot $t$ until the horizon is independent of the scheduling vector
$\vec{a}_{t+1}$ if the greedy policy is implemented in slots $\{t,t-1,\ldots, 1\}$, i.e.,  
\begin{eqnarray}
\sum_{k=t}^{1}\E_{\pi_k|{\vec{a}_{t+1},\pi_m}}R_k(\pi_k,\hat{a}_k)&=&\sum_{k=t}^{1}\E_{\pi_k|{\pi_m}}R_k(\pi_k,\hat{a}_k).\nonumber\\
\end{eqnarray}
Thus, if the greedy policy is optimal in slots $\{t,t-1,\ldots, 1\}$, then, it is also optimal in slot $t+1$.
Since $t$ is arbitrary and since the greedy policy is optimal at the horizon, by induction, the greedy
policy is optimal in every slot $\{m,m-1,\ldots, 1\}$. This establishes the proposition.
\end{proof}

{\bf Remarks:} When the Markov channels are negatively correlated, i.e., $p<r$ - the case of limited practical significance, using arguments similar to those in the preceding proof, we can show that the greedy policy is optimal when $N=2$, for any ARQ delay distribution. We record this below.
\begin{corollary}\label{necorr}
When the Markov channels are negatively correlated, i.e., $p<r$, and when $N=2$, the sum throughput, $\eta_{\textrm{sum}}(m,\{\textgoth{A}_{k}\}_{k=1}^{m})$, of the system is maximized by the greedy policy
$\{\widehat{\textgoth{A}}_k\}_{k=1}^{m}$ for any ARQ delay distribution.
\end{corollary}

A formal proof can be found in Appendix~\ref{app:cor2pt5}.

Returning to the original positive correlation setup, the arguments in the proof of Proposition~\ref{prop:myoopt} hold true even when the ARQ delay is not identically distributed across time. Thus, the greedy policy is optimal for $N=2$ even when the ARQ delay distribution is time-variant. Also, since $m$ is arbitrary, the greedy policy maximizes the sum throughput over an infinite horizon. We record this below.
\begin{corollary}
For $N=2$, the greedy policy is optimal when the performance metric is the sum throughput over an infinite
horizon, i.e.,
\begin{eqnarray}
\{\hat{\textgoth{A}}_{k}\}_{k\ge 1}&=&\arg\max_{\{\textgoth{A}_{k}\}_{k\ge 1}}\lim_{m\rightarrow
\infty}\frac{V_{m}(\pi,\{\textgoth{A}_{k}\}_{k\ge 1})}{m}
\end{eqnarray}
for any initial belief $\pi$.
\end{corollary}

% For the case when there is no ARQ delay, the current problem can be linked to a work on
% opportunistic spectrum access in cognitive radio systems in \cite{Zhao}. Here the authors show that
% the greedy policy is optimal for the general $N$ user system.

The optimality of the greedy policy does not extend to the case $N>2$. We record this in the following proposition.
\begin{proposition}\label{prop:counter}
%For $N>2$, $\exists$ a $N, P, P_D, m, \pi_m$
The greedy policy is not, in general, optimal when there are more than two users in the downlink.
\end{proposition}
% We first introduce the following lemma.
% \begin{lemma}
% The state evolution operator $T(.)$ satisfies the following properties:
% \begin{itemize}
% \item $T^u(p)>T^v(p)$ for $u<v$, $T^u(p)>T^v(r)$ for any $u,v$ and $T^u(p)>T^v(x)$ for $u<v$
% and any $x\in[0,1]$
% \item $T^v(x)>T^u(r)$ for $u<v$ and for any $\pi\in[0,1]$
% \item $T^v(r)>T^u(r)$ for $u<v$
% \end{itemize}
% \end{lemma}
% \begin{proof}
% Using $T^u(x)=x(p-r)^u+r\frac{(1-(p-r)^u)}{1-(p-r)}$ and $p>r$, the properties can be established
% with algebraic manipulations.
% \end{proof}
%\begin{proof}

\textit{Proof outline:} We establish the proposition using a counterexample with deterministic ARQ delay of $D=1$, i.e., $P_D(d=1)=1$, and arbitrary values of $N;N>2$ and $m;m>3$. We construct a variant of the greedy policy that schedules a non-greedy user in a specific time slot under a specific sample path of the past channel states observable by the scheduler. In the rest of the slots and under other realizations, the constructed policy performs greedy scheduling. We explicitly evaluate the difference in the rewards corresponding to the constructed policy and the greedy policy and show that, there exists system parameters such that the constructed policy has a reward strictly larger than the greedy policy. Thus the greedy policy is, in general, not optimal when $N>2$. A formal proof can be found in Appendix~\ref{app:propcounter}.

{\bf Remarks:} Note that, in contrast, it has been shown in \cite{Zhao} that the greedy policy is optimal for any number of users when the ARQ feedback is {\em instantaneous}, i.e., $D=0$. To summarize,  the optimality of the greedy policy vanishes 
\begin{itemize}
\item when the ARQ delay is increased from zero to higher values, with the number of users unconstrained, or 
\item when the number of users is increased from two to higher values, with the ARQ delay being random and unconstrained.
\end{itemize}
These observations point to the volatile nature of the underlying dynamics of the scheduling problem, with respect to the greedy policy optimality.

It would be interesting to see how the optimality properties of the greedy policy extend to more general channel models.  
Considering the multi-rate channels, i.e., when the number of states is greater than two, the special `toggle' structure that led to the optimality of the greedy policy in the ON-OFF channel vanishes. In fact, we have shown \cite{3state} that, even when the number of states is increased by 1, the general greedy policy optimality vanishes and the optimality can be shown to hold only under very restrictive conditions on the Markov channel statistics. Now, consider the case when the two-state Markov channels are non-identical across users. In this setup, we can show that the greedy policy is not, in general, optimal, even when the ARQ delay is instantaneous. We record this below.
\begin{proposition}\label{prop:4pt5}
The greedy policy is not, in general, optimal when the Markov channels are not identical across users, even when $N=2$ and the ARQ feedback is instantaneous.
\end{proposition}

The proposition is established using counterexamples. Proof is available in Appendix~\ref{app:prop4pt5}.

In summary, continuing our discussion before Proposition~\ref{prop:4pt5}, the optimality of the greedy policy vanishes even under minimal deviations from the original setup. These observations further indicate the volatile nature of the underlying scheduling problem dynamics. 

Returning to the original setup at hand, numerical results suggest that the greedy policy, despite being not optimal in general, has near optimal performance. We discuss this next.

\subsection{Performance Evaluation of the Greedy Policy}\label{sec:greedyperf}

\begin{table*}
\begin{center}
\renewcommand{\tabcolsep}{.4cm}
\renewcommand{\arraystretch}{1.65}
\begin{tabular}{|c|c|c|c|c|c|c|}
\hline
$N$ & Delay $=[P_D(0)\ldots P_D(d_{\max})]$  & $p$ & $r$ & $V_{\opt}$ & $V_{\greedy}$ & $\%$subopt \\
\hline
3 & $[0.8822~0.1178]$ & $0.9172$ & $0.2858$ & $6.0707$ & $6.0696$ & $0.0182~\%$\\
\hline
4 & $[0.5387~0.4613]$ & $0.9464$ & $0.1666$ & $5.9700$ & $5.9586$ & $0.1910~\%$\\
\hline
3 & $[0.5908~0.3959~0.0132]$ & $0.6619$ & $0.2389$ & $3.9933$ & $3.9914$ & $0.0476~\%$\\
\hline
4 & $[0.6647~0.1844~0.1510]$ & $0.9281$ & $0.2824$ & $5.8934$ & $5.8854$ & $0.1364~\%$\\
\hline
\end{tabular}
\end{center}
\caption{Comparison of the performance of the greedy policy with the optimal reward.}
\label{tab:1}
\end{table*}
%\vspace{-20pt}
%
%2 taps
\begin{table*}
\begin{center}
\renewcommand{\tabcolsep}{.25cm}
\renewcommand{\arraystretch}{1.75}
%\begin{tabular}{cc}
\begin{tabular}{|c|c|c|c|c|c|c|c|}
\hline
\multirow{6}{*}{N=10} & \multirow{2}{*}{Delay} & \multicolumn{3}{|c|}{$p=0.5848,~r=0.3509$} & \multicolumn{3}{|c|}{$p=0.6392,~r=0.2328 $}\\
\cline{3-8}
& & $V_{\genie}$ & $V_{\greedy}$ &  $\%$genie & $V_{\genie}$ & $V_{\greedy}$ & $\%$genie\\
\cline{2-8}
& $[0~ 1]$ & 5.3908 & 5.2912  & 1.8470$~\%$  & 5.5279 & 5.2067 & 5.8109$~\%$\\
\cline{2-8}
& $[\frac{1}{3}~ \frac{2}{3}]$ & 5.6547 & 5.4281 & 4.0072$~\%$  & 5.9195 & 5.4119 &  8.5741$~\%$\\
\cline{2-8}
& $[\frac{1}{2}~ \frac{1}{2}]$ & 5.7867 & 5.4987 & 4.9771$~\%$  & 6.1152 & 5.5208 & 9.7203$~\%$ \\
\cline{2-8}
& $[\frac{2}{3}~ \frac{1}{3}]$ & 5.9187 & 5.5712 &  5.8703$~\%$ & 6.3110 & 5.6353  & 10.7070$~\%$ \\
\hline\hline
\multirow{6}{*}{N=20} & \multirow{2}{*}{Delay} & \multicolumn{3}{|c|}{$p=0.9148,~r=0.4309$} & \multicolumn{3}{|c|}{$p=0.3079,~r=0.2517$}\\
\cline{3-8}
& & $V_{\genie}$ & $V_{\greedy}$ & $\%$genie & $V_{\genie}$ & $V_{\greedy}$ & $\%$genie\\
\cline{2-8}
& $[0~ 1]$ & 8.8565 & 8.8254 & 0.3504$~\%$ & 3.4487 & 3.4371 & 0.3368$~\%$\\
\cline{2-8}
& $[\frac{1}{3}~ \frac{2}{3}]$ & 8.9715 & 8.9291 & 0.4723$~\%$ & 3.5525 & 3.4661 & 2.4315$~\%$\\
\cline{2-8}
& $[\frac{1}{2}~ \frac{1}{2}]$ & 9.0290 & 8.9820 & 0.5203$~\%$ & 3.6043 & 3.4807 & 3.4300$~\%$\\
\cline{2-8}
& $[\frac{2}{3}~ \frac{1}{3}]$ & 9.0865 & 9.0357  & 0.5593$~\%$ & 3.6562 & 3.4955 & 4.3967$~\%$\\
\hline
\end{tabular}
\end{center}
\caption{Comparison of the performance of the greedy policy with the optimal reward in the genie-aided system. Maximum ARQ delay, $d_{\max}=1$.}
\label{tab:2}
\end{table*}
%
%3 taps
\begin{table*}
\begin{center}
\renewcommand{\tabcolsep}{.25cm}
\renewcommand{\arraystretch}{1.75}
%\begin{tabular}{cc}
\begin{tabular}{|c|c|c|c|c|c|c|c|}
\hline
\multirow{6}{*}{N=10} & \multirow{2}{*}{Delay} & \multicolumn{3}{|c|}{$p=0.2148,~r=0.1100$} & \multicolumn{3}{|c|}{$p=0.6863,~r=0.4136 $}\\
\cline{3-8}
& & $V_{\genie}$ & $V_{\greedy}$ &  $\%$genie & $V_{\genie}$ & $V_{\greedy}$ & $\%$genie \\
\cline{2-8}
& $[0~ 0~ 1]$ & 2.0196 & 2.0162 & 0.1716$~\%$  & 6.2768 & 6.2571  & 0.3131$~\%$\\
\cline{2-8}
& $[\frac{1}{6}~ \frac{1}{3}~ \frac{1}{2}]$ & 2.1261 & 2.0384 & 4.1241$~\%$  & 6.4895 & 6.3813 & 1.6663$~\%$\\
\cline{2-8}
& $[\frac{1}{3}~ \frac{1}{3}~ \frac{1}{3}]$ & 2.2152 & 2.0577 & 7.1089$~\%$ & 6.6375 & 6.4743 &2.4587$~\%$ \\
\cline{2-8}
& $[\frac{1}{2}~ \frac{1}{3}~ \frac{1}{6}]$ & 2.3018 & 2.0772 & 9.7568$~\%$ & 6.7764 & 6.5677 & 3.0792$~\%$ \\
\hline\hline
\multirow{6}{*}{N=20} & \multirow{2}{*}{Delay} & \multicolumn{3}{|c|}{$p=0.8822,~r=0.2816$} & \multicolumn{3}{|c|}{$p=0.7120,~r=0.5713 $}\\
\cline{3-8}
& & $V_{\genie}$ & $V_{\greedy}$ & $\%$genie & $V_{\genie}$ & $V_{\greedy}$ & $\%$genie \\
\cline{2-8}
& $[0~ 0~ 1]$ & 8.0485 & 7.9811 & 0.8376$~\%$  & 7.0084 & 7.0066 & 0.0251$~\%$ \\
\cline{2-8}
& $[\frac{1}{6}~ \frac{1}{3}~ \frac{1}{2}]$ & 8.3208 & 8.1880 &  1.5952$~\%$ & 7.0868 & 7.0585  & 0.3989$~\%$ \\
\cline{2-8}
& $[\frac{1}{3}~ \frac{1}{3}~ \frac{1}{3}]$ & 8.4754 & 8.3186 & 1.8493$~\%$ & 7.1495 & 7.1017 & 0.6675$~\%$ \\
\cline{2-8}
& $[\frac{1}{2}~ \frac{1}{3}~ \frac{1}{6}]$ & 8.6131 & 8.4490 &1.9057$~\%$ & 7.2099 & 7.1448 &  0.9018$~\%$ \\
\hline
\end{tabular}
\end{center}
\caption{Comparison of the performance of the greedy policy with the optimal reward in the genie-aided system. Maximum ARQ delay, $d_{\max}=2$.}
\label{tab:3}
\end{table*}
\begin{table*}
\begin{center}
\renewcommand{\tabcolsep}{.3cm}
\renewcommand{\arraystretch}{1.75}
\begin{tabular}{|c|c|c|c|c|c|c|c|c|c|}
\hline
$(p-r)$ & Delay  & $V_{\genie}$ & $V_{\greedy}$ & $\%$genie & $(p-r)$ & Delay  & $V_{\genie}$ & $V_{\greedy}$ & $\%$genie \\
\hline
\multirow{4}{*}{0.2} & $[0~0~1]$  &   $5.6342$  &  $5.6232$  &  $0.1953~\%$ & \multirow{4}{*}{0.8} & $[0~0~1]$  & $7.9848$ &  $7.7252$  &  $3.2520~\%$\\
\cline{2-5}\cline{7-10}
& $[\frac{1}{6}~ \frac{1}{3}~ \frac{1}{2}]$ & $5.8068$  &  $5.7105$  &  $1.6592~\%$ & & $[\frac{1}{6}~ \frac{1}{3}~ \frac{1}{2}]$ & $8.3585$ &  $8.0181$  &  $4.0726~\%$\\
\cline{2-5}\cline{7-10}
& $[\frac{1}{3}~ \frac{1}{3}~ \frac{1}{3}]$ & $5.9357$ &  $5.7797$ & $2.6283~\%$ & & $[\frac{1}{3}~ \frac{1}{3}~ \frac{1}{3}]$ &  $8.5551$ &  $8.1843$  &  $4.3347~\%$\\
\cline{2-5}\cline{7-10}
& $[\frac{1}{2}~ \frac{1}{3}~ \frac{1}{6}]$ & $6.0584$ &  $5.8494$  & $3.4499~\%$ & & $[\frac{1}{2}~ \frac{1}{3}~ \frac{1}{6}]$ & $8.7265$ & $8.3522$  &  $4.2890~\%$\\
\hline
\end{tabular}
\end{center}
\caption{Illustration of the effect of the Markov channel memory, $(p-r)$ on the reward functions. Maximum ARQ delay, $d_{\max}=2$.}
\label{tab:4}
\end{table*}

Table~\ref{tab:1} provides a sample of the net expected reward under the greedy policy ($V_{\greedy}$) in comparison with that of the optimal policy ($V_{\opt}$) when $N=3$, and when $N=4$, for horizon length $m=7$. The ARQ delay probability mass function is generated (uniform) randomly with the maximum delay $d_{\max}$ fixed first. The high values of the quantity $\%$subopt$=\frac{V_{\opt}-V_{\greedy}}{V_{\opt}}\times 100\%$ illustrates the near optimal performance of the greedy policy for the system parameters considered. Note that, the optimal reward, $V_{\opt}$, is evaluated by a brute-force search over the scheduling decisions in every slot $t\in\{m,m-1,\ldots,1\}$, that is prohibitively complex for larger values of $N$ and $m$. We, therefore, perform an indirect study of the greedy policy performance in Tables~\ref{tab:2}-\ref{tab:4}, that allows us to consider wider range of system parameters. We first define the genie-aided system as follows:  for any slot $k$, the feedback $f_k$ includes the channel state information, corresponding to slot $k$, of not only the scheduled user $a_k$ but also that of all the users in the system. Thus the optimal reward in the genie-aided system, $V_{\genie}$, is an upper bound to the optimal reward in the original system, $V_{\opt}$. Also, $V_{\genie}$ can be evaluated using closed-form expressions, with complexity much lower than that of $V_{\opt}$. We will discuss the evaluation of $V_{\genie}$ in the context of the genie-aided system sum capacity in Section~\ref{sec:sumC}.

In Table~\ref{tab:2}, with the maximum ARQ delay $d_{\max}=1$, the net expected reward under the greedy policy is compared with $V_{\genie}$ when $N=10$ and when $N=20$, for randomly generated values of $p$ and $r$. The length of the horizon is fixed at $m=10$. The probability mass function of the ARQ delay, denoted by `Delay' in the table, is controlled to have a weakening `tail' from $[0~1]$ to $[\frac{2}{3} \frac{1}{3}]$. The quantity $\%$genie$=\frac{V_{\genie}-V_{\greedy}}{V_{\genie}}\times 100\%$ is an upper bound to the quantity $\%$subopt introduced earlier. Table~\ref{tab:3} is similarly constructed with the maximum ARQ delay $d_{\max}=2$. In both Tables~\ref{tab:2} and \ref{tab:3}, we see that  $\%$genie is predominantly low-valued, suggesting that the greedy policy has near optimal performance. Also, note that, as the tail of the ARQ delay mass function weakens, both $V_{\genie}$ and $V_{\greedy}$ increase. This is expected since, with a weakening tail, the ARQ feedback is stochastically more `fresh', thereby facilitating better informed scheduling decisions and higher rewards in both genie-aided and original systems. Also note that, as the tail weakens, the gap between the \textit{optimal} rewards in the genie-aided system and the original system can be expected to increase, since the gap between the information content of the full feedback (genie-aided system) and the ARQ feedback increases with a weakening tail. Thus, the relatively high values of $\%$genie corresponding to weaker delay tails, could be due to an inherent system level gap between the genie-aided and the original systems, and need not necessarily be a pointer to the greedy policy performance. The last statement is further strengthened by the fact that the greedy policy is optimal when the ARQ delay tail is at the weakest, i.e., when the feedback is \cite{Zhao}. 

In Table~\ref{tab:4}, we study the effect of the Markov channel memory, defined as $(p-r)$, on the reward functions. With $d_{\max}=2$, $m=10$ and $N=20$, we consider two extreme values of the channel memory, i.e., $(p-r)=0.2$ and $(p-r)=0.8$. In both cases of channel memory, we have fixed the steady state probability of the ON state to be $\pi_{ss}=0.5$, by fixing $p+r=1$. This essentially provides a degree of fairness when comparing these two cases. Note that, for a fixed delay statistic, the rewards $V_{\genie}$ and $V_{\greedy}$ increase with increase in the channel memory. This is due to an increase in the value of the feedback, as the channel memory increases. Also, we see an increase in the value of $\%$genie as the memory increases. This points to two underlying phenomena: 1) An increase in the inherent sub-optimality associated with greedy scheduling as the channel memory increases 2) Similar to the case of weakening delay tail, an increase in the channel memory results in an increase in the system level gap between the genie-aided and the original systems, by way of an increase in the gap between the information content of full feedback (genie-aided system) and the ARQ feedback. 

Summarizing, Tables~\ref{tab:1}-\ref{tab:4} suggest that the greedy policy has near optimal performance for a wide range of system parameters and that the ARQ delay profile and the channel memory affect the reward values in ways that can be explained intuitively. In addition, note that $\%$genie is also an upper bound to the quantity  $\frac{V_{\genie}-V_{\opt}}{V_{\genie}}\times 100\%$. Thus the low values of $\%$genie provide the following larger message: using only the 1-bit ARQ feedback for opportunistic scheduling is associated with system level performance comparable to the case when feedback is available from \textit{all} the users.

%
%
%
%
%
%
% \putFrag{comparisonplot}{Average total reward of the greedy policy in
% comparison with system-level performance limits. System parameters used: plot (A) $N=3$, $p=0.4070$,
% $r=0.1999$, $P_D(0)=0.3379,P_D(1)=0.5666,P_D(2)=0.0954$, $\pi_m=[0.7487~0.8256~0.7900]$, (B) $N=3$,
% $p=0.9930$, $r=0.1267$, $P_D(0)=0.8855,P_D(1)=0.1145$, $\pi_m=[0.3631~0.2662~0.3857]$, (C) $N=3$,
% $p=0.9694$, $r=0.1556$, $P_D(0)=0,P_D(1)=1$, $\pi_m=[0.1207~0.1962~0.1791]$, (D) $N=3$, $p=0.7965$,
% $r=0.1365$, $P_D(0)=0,P_D(1)=0,P_D(2)=1$, $\pi_m=[0.1351~0.2523~0.2410]$}{18}{}

\subsection{Structure of the Greedy Policy}\label{sec:struc}
Motivated by the near optimal performance of the greedy policy, we proceed to study its structure,
which turns out to be very amenable for practical implementation. We begin by defining the following
quantity:\\
\textit{Schedule order vector}, $O_{t}$, in slot $t$: The user indices in
decreasing order of $\pi_{t}(i)$, i.e.,
\begin{eqnarray}
O_{t}(1)&=&\arg\max_{i}\pi_{t}(i)\nonumber\\
\vdots\nonumber\\
O_{t}(N)&=&\arg\min_{i}\pi_{t}(i).\nonumber
\end{eqnarray}
Thus, the greedy decision in slot $t$ is $\hat{a}_{t}=O_{t}(1)$.

Now, in any slot $t\le m$, any user $i$ falls under one of the following two cases:
\begin{itemize}
\item[]\hspace{-12pt}1) The scheduler has received at least one ARQ feedback from user $i$ by the beginning of slot $t$. Let
$k_i$, for $m\ge k_i >t$, be the latest slot for which the ARQ feedback from user $i$ is available at the
scheduler. Since the channel is first-order Markovian, the belief value of the channel of user $i$
in the current slot $t$ is dependent only on the feedback $f_{k_i}$ and $k_i$. The belief value is
given by
\begin{eqnarray}\label{eq:case1}
\pi_t(i)=\begin{cases}
T^{k_i-t-1}(p) & \mbox{if } f_{k_i}=1 \\
T^{k_i-t-1}(r) & \mbox{if } f_{k_i}=0. \\
\end{cases}
\end{eqnarray}

\item[]\hspace{-12pt}2) The scheduler does not have any ARQ feedback from user $i$ by the beginning of slot $t$.
In this case
\begin{eqnarray}\label{eq:case2}
\pi_t(i)=T^{(m-t)}\pi_m(i).
\end{eqnarray}
Recall that $\pi_m(i)$ is the initial belief value of the
channel of user $i$ when the scheduling process started at slot $m$.
\end{itemize}
At slot $t$, let $\mathcal{A}_t$ denote the set of users, $i$, whose latest feedback,
$f_{k_i}$, is an ACK.
Let $\mathcal{N}_t$ denote the set of users, $j$, whose latest feedback, $f_{k_j}$, is a NACK. Let
the users from whom the scheduler has not yet received any feedback constitute set $\mathcal{X}_t$.
From (\ref{eq:case1}) and (\ref{eq:case2}), using Lemma~\ref{le:Tprop}, the
greedy decision in slot $t$ can be written as
\begin{eqnarray}\label{eq:3cases}
\hat{a}_{t}=\begin{cases}
\arg\min_{i\in\mathcal{A}_t}k_i & \mbox{if } \mathcal{A}_t\neq \emptyset \\
\arg\max_{i\in\mathcal{X}_t}\pi_m(i) & \mbox{if } \mathcal{A}_t = \emptyset ~\mbox{and}~
\mathcal{X}_{t}\neq \emptyset \\
\arg\max_{i\in\mathcal{N}_t}k_i & \mbox{if } \mathcal{A}_t = \emptyset ~\mbox{and}~
\mathcal{X}_{t}= \emptyset. \\
\end{cases}
\end{eqnarray}
Now, for ease of implementation, we visualize the sets $\mathcal{A}_t$, $\mathcal{X}_t$ and $\mathcal{N}_t$ as queues
with elements ordered in the following specific ways: Let $\mathcal{A}_t(i)$ denote the $i^{th}$ element
of queue $\mathcal{A}_t$ and the elements be ordered such that
$k_{\mathcal{A}_t(1)}<k_{\mathcal{A}_t(2)}\ldots < k_{\mathcal{A}_t(n(A_t))}$, where $n(A)$ denotes
the cardinality of set $A$. Note that the user that gave
an ACK from the most recent slot lies at the head of queue $\mathcal{A}_t$. The elements of
$\mathcal{X}_t$ are ordered such that $\pi_m(\mathcal{X}_t(1))\ge \pi_m(\mathcal{X}_t(2))\ldots
\ge \pi_m(\mathcal{X}_t(n(X_t)))$. The elements of $\mathcal{N}_t$ satisfy
$k_{\mathcal{N}_t(1)}>k_{\mathcal{N}_t(2)}\ldots > k_{\mathcal{N}_t(n(N_t))}$, i.e., the user with the
oldest NACK feedback lies on top of queue $\mathcal{N}_t$. Define a combined queue constructed by
concatenating the queues $\mathcal{A}_t$, $\mathcal{X}_t$ and $\mathcal{N}_t$ in that order. From (\ref{eq:case1}) and (\ref{eq:case2}), using Lemma~\ref{le:Tprop}, we see that the users in the combined queue are arranged in decreasing order (top-down)
of belief values with the top-most user being the greedy decision in slot $t$. Thus the combined queue
is, in fact, the schedule order vector $O_t$.
\begin{figure*}
\centering
\psfrag{Begin}[cc]{\tiny{Beginning of slot $t$}}
\psfrag{Top}[cc]{\tiny{Schedule the user on top ($\hat{a}_t$)}}
\psfrag{KiAt}[cc]{\tiny{~~~~~$k_i$, $i\in A_t$}}
\psfrag{increases}[cc]{\tiny{~~~~increases}}
\psfrag{setAt}[cc]{\tiny{set $A_t$}}
\psfrag{pimXt}[cc]{\tiny{   $\pi_m(i)$, $i \in X_t$}}
\psfrag{decreases}[cc]{\tiny{    decreases}}
\psfrag{setXt}[cc]{\tiny{set $X_t$}}
\psfrag{KiNt}[cc]{\tiny{  $k_i$, $i\in N_t$}}
\psfrag{setNt}[cc]{\tiny{set $N_t$}}
\psfrag{Ap}[cc]{\tiny{$A_+$}}
\psfrag{Am}[cc]{\tiny{$A_-$}}
\psfrag{Np}[cc]{\tiny{$N_+$}}
\psfrag{Nm}[cc]{\tiny{$N_-$}}
\psfrag{Ot}[cc]{\tiny{$O_t$}}
\psfrag{Rma}[cc]{$\begin{array}{c}\vspace{-5pt}\textrm{\tiny{Remove $a$ from its}}\\\vspace{-5pt}\textrm{\tiny{current
position.}}\\\vspace{-5pt}\textrm{\tiny{Insert here such that}}\\\textrm{\tiny{$k_{A_+}<k_a<k_{A_-}$}}\end{array}$}
\psfrag{Rmb}[cc]{$\begin{array}{c}\vspace{-5pt}\textrm{\tiny{Remove $a$ from its}}\\\vspace{-5pt}\textrm{\tiny{current
position.}}\\\vspace{-5pt}\textrm{\tiny{Insert here such that}}\\\textrm{\tiny{$k_{N_+}>k_a>k_{N_-}$}}\end{array}$}
\psfrag{E}[cc]{$\begin{array}{c}\vspace{-5pt}\textrm{\tiny{End of slot
$t$.}}\\\vspace{-5pt}\textrm{\tiny{$F_t$ received}}\end{array}$}
\psfrag{F}[cc]{$\begin{array}{c}\vspace{-5pt}\textrm{\tiny{For every user
$a$}}\\\vspace{-5pt}\textrm{\tiny{whose ARQ feedback}}\\\vspace{-5pt}\textrm{\tiny{is contained in
$F_t$}}\end{array}$}
\psfrag{G}[cc]{$\begin{array}{c}\vspace{-5pt}\textrm{\tiny{If}}\\\vspace{-5pt}\textrm{\tiny{this is
the latest}}\\\vspace{-5pt}\textrm{\tiny{feedback for}}\\\vspace{-5pt}\textrm{\tiny{user $a$}}\end{array}$}
\psfrag{H}[cc]{$\begin{array}{c}\vspace{-5pt}\textrm{\tiny{Leave
$a$}}\\\vspace{-5pt}\textrm{\tiny{in its}}\\\vspace{-5pt}\textrm{\tiny{current position}}\end{array}$}
\psfrag{I}[cc]{$\begin{array}{c}\vspace{-5pt}\textrm{\tiny{If}}\\\vspace{-5pt}\textrm{\tiny{$f_{k_a}=1$}}\end{array}$}
\psfrag{Y}[cc]{\tiny{YES}}
\psfrag{N}[cc]{\tiny{NO}}
\includegraphics[width=.7\textwidth]{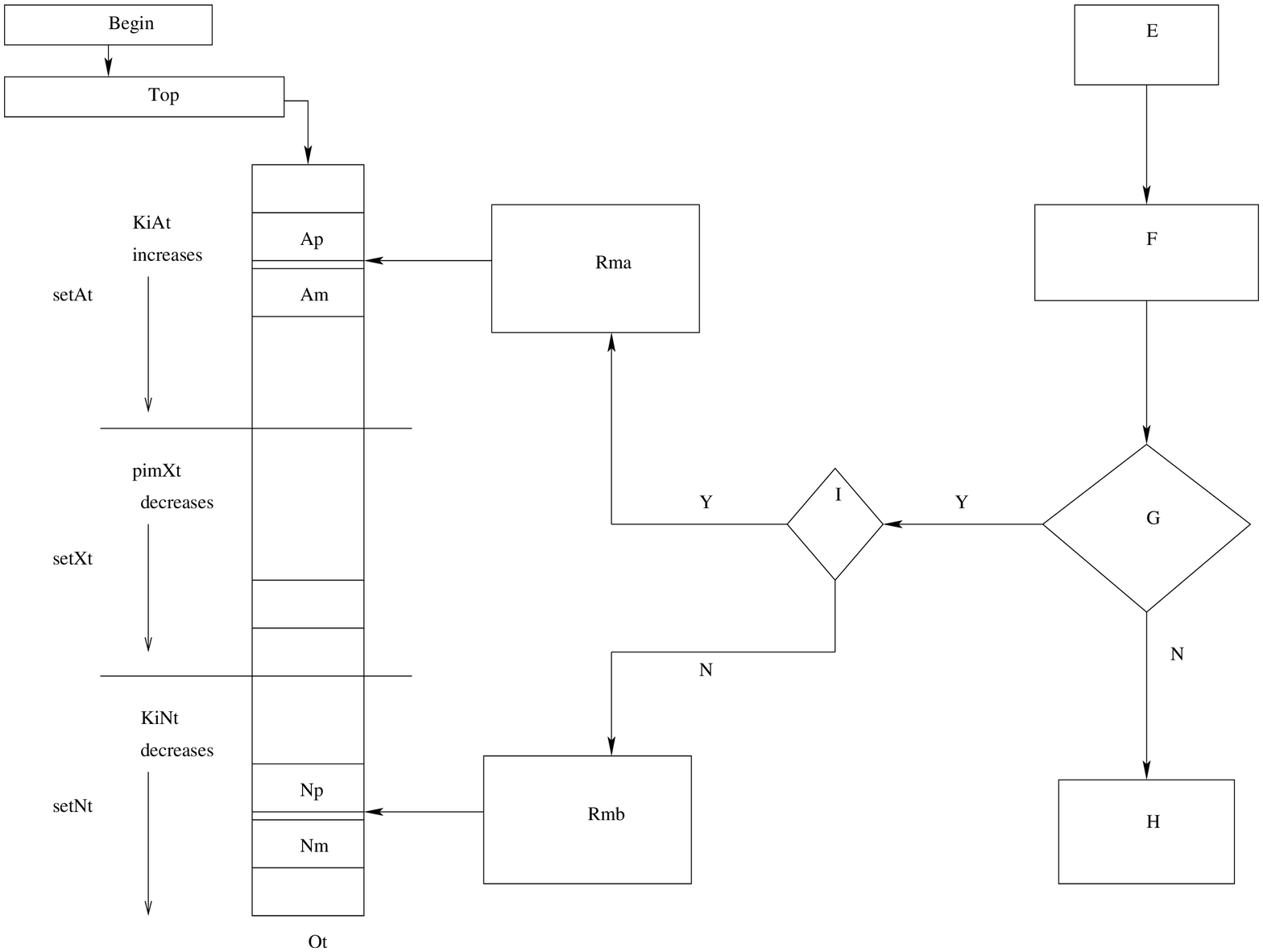}
\caption{Greedy policy implementation under random ARQ delay.} \label{fig:Fig0}
\end{figure*}

We now discuss the evolution of the schedule order vector.
For every user $a$ whose ARQ feedback is contained in $F_t$, implement the following procedure: Let
$t_a$ indicate the originating slot for the ARQ feedback from user $a$ contained in $F_t$.
Now, if $t_a$ is the latest slot from which the ARQ feedback of user $a$ is available at the
scheduler, then $k_a=t_a$. The new schedule
order vector $O_{t-1}$ is formed by removing user $a$ from its current position (in $O_t$) and
placing it in the sub-queue
$\mathcal{A}_{t-1}$ (if $f_{k_a}=1$) or in the sub-queue $\mathcal{N}_{t-1}$ (if $f_{k_a}=0$) at an
appropriate location (so that the ordering based on $k_i$ is not violated). If $t_a\neq k_a$, i.e.,
$t_a$ is not the latest slot, then user $a$ is not moved. Similarly, users whose ARQ feedback are
not contained in $F_t$ are not moved. The last two statements are direct consequences of the following
facts:
\begin{itemize}
\item For an user $a$ whose ARQ feedback is contained in $F_t$ but is not the latest feedback from
that user, the belief value evolves as $\pi_{t-1}(a)=T(\pi_{t}(a))$. Similarly, for an user $b$ whose ARQ feedback
is \textit{not} contained in $F_t$, the belief value evolves as $\pi_{t-1}(b)=T(\pi_{t}(b))$. Both these
cases were discussed in Section~\ref{subsec:prob defn}.
\item From Lemma~\ref{le:Tprop}, if $x\ge y$, then $T(x)\ge T(y)$.
\end{itemize}
Now, at slot $t-1$, the user on top of $O_{t-1}$ is the greedy decision. Thus the greedy decision in any slot is determined
by the latest ARQ feedback and the corresponding originating slot index of all the users in the system. Note
that this implementation does not require the Markov channel statistics (other than the knowledge that
$p>r$) and the statistics of the ARQ feedback delay. An illustration of the greedy policy
implementation is provided in \figref{Fig0}. 

\begin{figure}
\centering
\psfrag{A}[cc]{\tiny{Beginning of slot $t$}}
\psfrag{B}[cc]{\tiny{Schedule the user on top ($\hat{a}_t$)}}
\psfrag{C}[cc]{$\begin{array}{c}\vspace{-5pt}\textrm{\tiny{Remove $\hat{a}_{t+d}$ from
its}}\\\vspace{-5pt}\textrm{\tiny{current position.}}\\
\vspace{-5pt}\textrm{\tiny{Place on top}}\end{array}$}
\psfrag{D}[cc]{$\begin{array}{c}\vspace{-5pt}\textrm{\tiny{Remove $\hat{a}_{t+d}$ from
its}}\\\vspace{-5pt}\textrm{\tiny{current position.}}\\
\vspace{-5pt}\textrm{\tiny{Place at the bottom}}\end{array}$}
\psfrag{E}[cc]{\tiny{End of slot $t$}}
%
% \psfrag{E}[cc]{$\begin{array}{c}\vspace{-5pt}\textrm{\tiny{End of slot
% $t$.}}\\\vspace{-5pt}\textrm{\tiny{$F_t$ received}}\end{array}$}
\psfrag{F}[cc]{$\begin{array}{c}\vspace{-5pt}\textrm{\tiny{$F_t=f_{t+d}$
received}}\\\vspace{-5pt}\textrm{\tiny{from user $\hat{a}_{t+d}$}}\end{array}$}
\psfrag{G}[cc]{$\begin{array}{c}\vspace{-5pt}\textrm{\tiny{If}}\\\vspace{-5pt}\textrm{\tiny{$F_t=1$}}\end{array}$}
\psfrag{Y}[cc]{\tiny{YES}}
\psfrag{N}[cc]{\tiny{NO}}
\psfrag{Ot}[cc]{\tiny{$O_t$}}
\includegraphics[width=0.45\textwidth]{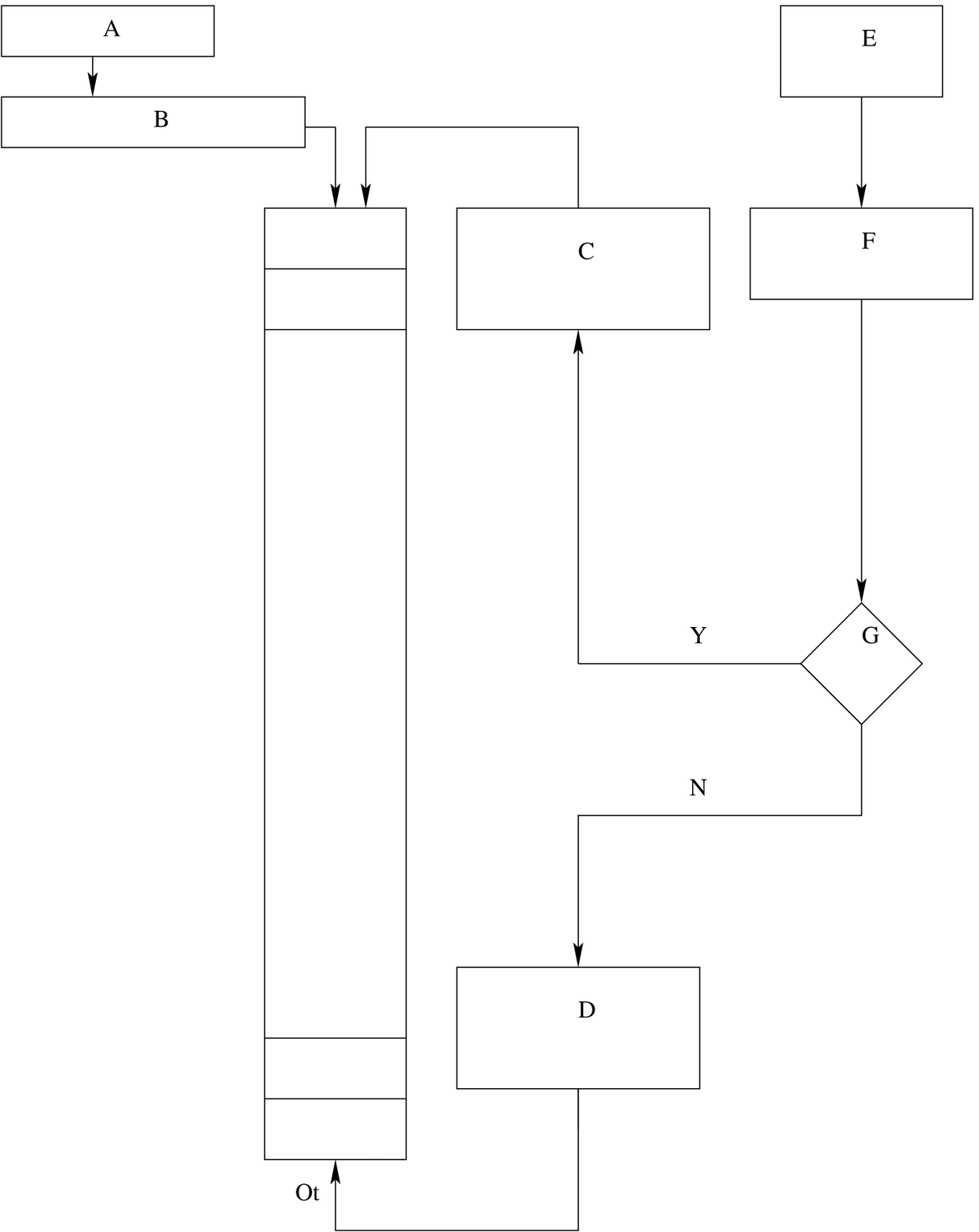}
\caption{Greedy policy implementation under deterministically delayed ARQ, i.e., $D=d$.} \label{fig:Fig1}
\end{figure}

\begin{figure}
\centering
\psfrag{A}[cc]{\tiny{Beginning of slot $t$}}
\psfrag{B}[cc]{\tiny{Schedule the user on top ($\hat{a}_t$)}}
%\psfrag{C}[cc]{\tiny{Leave $a_{t}$ on top}}
\psfrag{C}[cc]{$\begin{array}{c}\vspace{-5pt}\textrm{\tiny{$\hat{a}_{t}$ remains on
top.}}\\\vspace{-5pt}\textrm{\tiny{It will be rescheduled}}\\
\vspace{-5pt}\textrm{\tiny{in the next slot}}\end{array}$}
\psfrag{D}[cc]{\tiny{Place $\hat{a}_t$ at the bottom}}
% 
% \psfrag{D}[cc]{$\begin{array}{c}\vspace{-5pt}\textrm{\tiny{Remove $a_{t}$ from
% the}}\\\vspace{-5pt}\textrm{\tiny{top of $O_t$.}}\\
% \vspace{-5pt}\textrm{\tiny{Place at the bottom}}\end{array}$}
\psfrag{E}[cc]{\tiny{End of slot $t$}}
\psfrag{F}[cc]{$\begin{array}{c}\vspace{-5pt}\textrm{\tiny{$F_t=f_t$
received}}\\\vspace{-5pt}\textrm{\tiny{from user $\hat{a}_{t}$}}\end{array}$}
\psfrag{G}[cc]{$\begin{array}{c}\vspace{-5pt}\textrm{\tiny{If}}\\\vspace{-5pt}\textrm{\tiny{$F_t=1$}}\end{array}$}
\psfrag{Y}[cc]{\tiny{YES}}
\psfrag{N}[cc]{\tiny{NO}}
\psfrag{Ot}[cc]{\tiny{$O_t$}}
\includegraphics[width=0.45\textwidth]{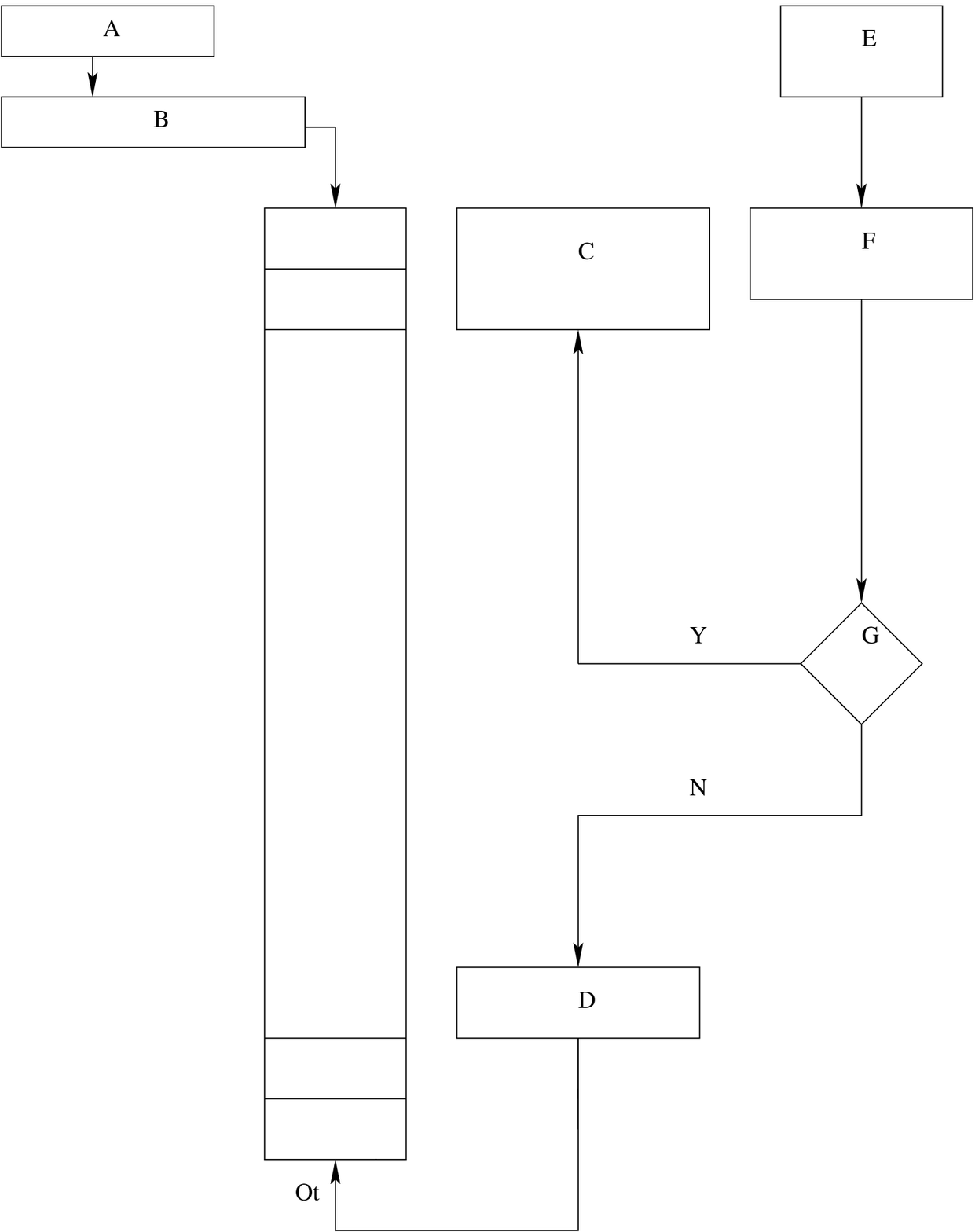}
\caption{Greedy policy implementation under instantaneous (end of slot) ARQ, i.e., $D=0$.}\label{fig:Fig02}
\end{figure}

For the special case of deterministic ARQ feedback delay $D=d$, the evolution from $O_t$ to $O_{t-1}$
is greatly simplified as follows. At the end of slot $t$, since $D=d$, $F_{t}$ contains feedback only from the
user scheduled in slot $t+d$, i.e., user $\hat{a}_{t+d}$. Thus $F_t=f_{t+d}$. The feedback bits
$f_m,f_{m-1},\ldots,f_{t+d+1}$ from users $\hat{a}_m,\hat{a}_{m-1},\ldots,\hat{a}_{t+d+1}$ have already arrived at the end
of slots $m-d,m-1-d,\ldots,t+1$ and the feedback from users $\hat{a}_{t+d-1},\hat{a}_{t+d-2},\ldots $ are yet to arrive.
% $a$, then, since the delay is deterministic, this must be the latest feedback from $a$, i.e.,
% $k_a=t+d$, and this must be the latest feedback from any user, i.e., $k_a=\min_{i\neq a}k_i$, since any feedback from slots
% $t+d-1,\ldots, t$ has not arrived yet. 
Thus $F_t=f_{t+d}$ from user $\hat{a}_{t+d}$ is the latest feedback available from \textit{any} user. Thus,
recalling the ordering rules for $\mathcal{A}_{t-1}$ and $\mathcal{N}_{t-1}$, if
$F_{t}=1$, user $\hat{a}_{t+d}$ is removed from its current position and placed on top in the updated schedule order vector, i.e.,
$O_{t-1}=[\hat{a}_{t+d}~~~O_{t}-\hat{a}_{t+d}]$, \footnote{If $Z=[z_1~z_2~z_3]$ then
$Z-z_2:=[z_1~z_3]$ and hence $[z_2~~Z-z_2]=[z_2~z_1~z_3]$} (user $\hat{a}_{t+d}$ becomes the greedy
decision in slot $t-1$). If $F_t=0$, $\hat{a}_{t+d}$ is placed at the bottom, i.e.,
$O_{t-1}=[O_{t}-\hat{a}_{t+d}~~~\hat{a}_{t+d}]$. When there is no ARQ delay ($D=d=0$), the
implementation becomes even simpler: on receiving an ACK, $O_{t-1}=O_t$, and on NACK,
$O_{t-1}=[O_t-O_t(1)~~~O_t(1)]$, since $\hat{a}_{t+d}=\hat{a}_t=O_t(1)$. This results in a simple round robin implementation of the greedy
policy as discussed in \cite{Murugesan,Zhao}. \figref{Fig1} and \figref{Fig02} illustrate the greedy policy implementation in
the deterministically delayed ARQ and instantaneous ARQ systems, respectively.

%\putFrag{Fig0}{Greedy policy implementation structure under random ARQ delay}{15}{}
%\putFrag{Fig1}{Greedy policy implementation structure under deterministic ARQ delay, $d$}{10}{}

%\putFrag{Fig2}{Greedy policy structure under instantaneous (end of slot) ARQ feedback }{12}{}

\section{On Downlink Sum Capacity and Capacity Region}\label{sec:limits}

We now proceed to study the fundamental limits on the downlink system performance --- the sum capacity and
the capacity region.
\subsection{Sum Capacity of the Downlink}\label{sec:sumC}
The sum capacity of the downlink is defined as the maximum sum throughput over an infinite
horizon with steady state initial conditions. Formally, with $N$ users in the system,
\begin{eqnarray}
C_{\summ}(N)&=&\max_{\{\textgoth{A}_{k}\}_{k\ge 1}}\lim_{m\rightarrow
\infty}\frac{V_{m}(\pi_{ss},\{\textgoth{A}_{k}\}_{k\ge 1})}{m},
\end{eqnarray}
where $\forall i\in\{1,\ldots, N\}$, $\pi_{ss}(i)=p_s$, the steady state probability of the Markov channel.
We now proceed to derive $p_s$. The Markov chain transition matrix
$P=\begin{bmatrix}
p&1-p\\
r&1-r\end{bmatrix}$
can be expressed as $P=U\Lambda V$, where

\begin{eqnarray}
U&=&\begin{bmatrix}
1&1\\
1&\frac{-r}{1-p}\end{bmatrix}\nonumber\\
\Lambda&=&\begin{bmatrix}
1&0\\
0&p-r\end{bmatrix}\nonumber\\
V&=&\frac{1}{1+\frac{r}{1-p}}\begin{bmatrix}
\frac{r}{1-p}&1\\
1&-1\end{bmatrix},\nonumber
\end{eqnarray}
with $VU=\begin{bmatrix}1&0\\0&1\end{bmatrix}$. Assuming\footnote{$p+(1-r)=2$ leads to
$P=\begin{bmatrix}1&0\\0&1\end{bmatrix}$, a trivial case with no steady state.} $p+(1-r)<2$,
\begin{eqnarray}
\lim_{n\rightarrow\infty}P^{n}&=&\begin{bmatrix}
\frac{r}{1-(p-r)}&1-\frac{r}{1-(p-r)}\\
\frac{r}{1-(p-r)}&1-\frac{r}{1-(p-r)}\end{bmatrix}\nonumber\\
\Rightarrow p_s&=&\frac{r}{1-(p-r)}.\nonumber
\end{eqnarray}
%
% With the greedy policy established as the sum-throughput maximizing scheduling policy for $N=2$
% users, the sum capacity of the system is given by the sum throughput under the greedy policy.
Recall, from Section~\ref{sec:greedyperf}, the definition of the genie-aided system: In any slot $k$, the feedback $f_k$ contains the channel state information, corresponding to slot $k$, of not only the scheduled user but also that
of all the users in the system. Also, the delay profile from the original system is retained in the genie-aided system,
i.e., the \textit{cumulative} feedback $f_k$ arrive at the scheduler with delay $D(a_k,k)$ that
is \emph{i.i.d} across scheduling choice $a_k$ and originating slot $k$ with the probability mass function $P_D(d)$. 
Thus, thanks to the cumulative nature of the feedback, the scheduling decision in the current slot does not affect the information available for scheduling in future slots. Hence, the greedy policy is optimal in the genie-aided system. With this insight, we now report our result on the sum capacity of the original downlink with two users.

\begin{proposition}\label{prop:capacity}
When $N=2$, the sum capacity of the Markov-modeled downlink with randomly delayed ARQ equals that of
the genie-aided system. This sum capacity equals
\begin{eqnarray}\label{eq:csum}
\lefteqn{C_{\summ}(N=2)}\nonumber\\
&=&\sum_{l=0}^{\infty}\Big[p_{s}T^l(p)+(1-p_s)p_s\Big]P(D\le
l)\prod_{d=0}^{l-1}P(D>d).\nonumber\\
\end{eqnarray}
Furthermore, the greedy policy achieves this sum capacity.
\end{proposition}

\begin{proof}
We first focus on the sum capacity of the genie-aided system, i.e., the sum throughput of
the greedy policy in the genie-aided system. Recall, from Section~\ref{sec:optproof}, the quantity $L$ -- the measure of freshness of the latest ARQ
feedback. We defined $L$ such that $L=l\Rightarrow$ the latest feedback is $l+1$ slots old. We
extend the meaning of $L$ to the genie-aided system. Due to the first order Markovian nature of the
channels, in the genie-aided system, conditioned on the latest feedback, $f_{t+l+1}$ (with $t$ denoting the current slot), the belief values (and hence the greedy scheduling decision) in the current slot are independent of the feedback from previous slots, i.e., $f_{k,k>t+l+1}$. Thus, with
$R_{\genie}^{\greedy}(l,N)$ denoting the conditional (conditioned on $L=l$) immediate reward
corresponding to the greedy policy, in the $N$-user genie-aided system with steady state initial conditions, the sum capacity of the genie-aided system can be written as
\begin{eqnarray}\label{eq:csummgenie}
C_{\summ}^{\genie}(N)&=&\E_l\big[ R_{\genie}^{\greedy}(l,N)\big].
\end{eqnarray}
%
% , the sum rate achieved by the
% greedy policy in the genie-aided system can be written as
% % \begin{eqnarray}
% $E_L( R_{\summ}^{\greedy, \genie}(l))$.
% % \end{eqnarray}
% % where $R_{\summ}^{\greedy, \genie}(l)$ is the sum rate of the greedy policy in the genie-aided system when the latest feedback
% % (from all the users) is $l+1$ slots old, i.e., the latest channel slot for
% % which channel state information is available is $l+1$ slots before the current slot. Recall that the quantity $l$ was used
% %in the optimality proof of greedy policy in section 3.1.
% Recall that $L=l\Rightarrow$ the latest feedback(s) is $l+1$ slots old. The belief value (in the current slot) of
% an user with an ON channel $l+1$ slots earlier ($T^l(p)$) is higher than the belief value of an user
% with an OFF channel $l+1$ slots earlier ($T^l(r)$). Also,
% %
% Thus, conditioned on $L=l$, the sum capacity of the
% genie-aided system, $r_{\summ}^{\opt,\genie}(l)$, equals the sum rate of the greedy policy in the
% genie-aided system conditioned on $L=l$, i.e., $r_{\summ}^{\greedy, \genie}(l)$.
We now evaluate $R_{\genie}^{\greedy}(l,N)$. From Lemma~\ref{le:Tprop}, the belief value (in the current slot) of
an user with an ON channel $l+1$ slots earlier, i.e., $T^l(p)$, is higher than the belief value of an user
with an OFF channel $l+1$ slots earlier, i.e., $T^l(r)$. Thus, in steady state,
% in slot $t$, the greedy policy schedules
% the user that was ON $l+1$ slots earlier, upon existence of such an user.
\begin{eqnarray}\label{eq:ropt}
\lefteqn{R_{\genie}^{\greedy}(l,N)}\nonumber\\
&=&P(\textrm{at least one of the $N$ users has an ON channel in }\nonumber\\
&&\hspace{143pt}\textrm{steady state})T^l(p)\nonumber\\
& &\hspace{-7pt}+P(\textrm{all users have OFF channels in steady state})T^l(r)\nonumber\\
&=&(1-(1-p_s)^N)T^l(p)+(1-p_s)^NT^l(r).
\end{eqnarray}
By explicitly including the probability mass function of the quantity $l$ as a function of the ARQ delay statistics, from (\ref{eq:csummgenie}) and (\ref{eq:ropt}), we have
\begin{eqnarray}\label{eq:csumgenie}
\lefteqn{C_{\summ}^{\genie}(N)}\nonumber\\
&=&\E_l R_{\genie}^{\greedy}(l,N)\nonumber\\
&=&\sum_{l=0}^{\infty}\Big[(1-(1-p_s)^N)T^l(p)+(1-p_s)^NT^l(r)\Big]\times\nonumber\\
&&\hspace{50pt}P(D\le l)\prod_{d=0}^{l-1}P(D>d).
\end{eqnarray}
When $N=2$, with minor algebraic manipulations, we have
\begin{eqnarray}\label{eq:csumgenie2}
\lefteqn{C_{\summ}^{\genie}(2)}\nonumber\\
&=&\sum_{l=0}^{\infty}\Big[p_{s}T^l(p)+(1-p_s)p_s\Big]P(D\le
l)\prod_{d=0}^{l-1}P(D>d).\nonumber\\
\end{eqnarray}

We now proceed to prove that the sum throughput of the greedy policy in the original
system equals that of the greedy policy in the genie-aided system when $N=2$.
% Consider the scheduling problem for the original system in slot
% $t$ under the greedy policy.
We established in the course of the proof of
Proposition~\ref{prop:myoopt} that, in the original system with $N=2$, conditioned on $L=l$, the
greedy decision in the current slot $t$ is solely determined by the ARQ feedback from slot $t+l+1$
with the following decision rule: When the user scheduled in slot $t+l+1$, i.e., $a_{t+l+1}$,
sends back an ACK, that user is scheduled in the current slot $t$, i.e., $\hat{a}_{t}=a_{t+l+1}$. Otherwise, the
other user is scheduled in slot $t$. We can interpret this decision logic of the greedy policy as
below:\\

\noindent\textit{When at least one of
the users had an ON channel in slot $t+l+1$, that user\footnote{User $a_{t+l+1}$ is given
higher priority if both channels were ON.} is identified for scheduling
in the current slot $t$, leading to an expected current reward of $T^l(p)$. Reward
$T^l(r)$ is accrued only when both the channels were in the OFF state in slot $t+l+1$.}\\

\noindent Note that the decision rule and the accrued immediate rewards corresponding to the greedy
policy in the original system are the same as that of the greedy policy in the genie-aided system.
Thus, in the original system, under the greedy policy,
no improvement in the immediate reward can be achieved even if the channel states
of both the users in slot $t+l+1$ are available at the scheduler in
slot $t$. This, along with the fact that both the systems have the same delay profile, establishes the equivalence between the original and the
genie-aided systems, when $N=2$, in terms of the sum throughput achieved by the greedy policy. We have already
proved the sum throughput optimality of the greedy policy in the original system when $N=2$
(Proposition~\ref{prop:myoopt}) and in the genie-aided system for a general value of $N$. Thus the sum capacity of the original system for
$N=2$ is given by $C_{\summ}^{\genie}(2)$ in (\ref{eq:csumgenie2}).
% We now proceed to prove that~(\ref{eq:sumc}) is the sum capacity of the genie-aided system as
% well by examining the sum throughput optimality of the greedy
% policy in the genie-aided system. For any control interval $m$, we rewrite the net expected reward
% from~(\ref{eq:Vm}) for the genie
% aided system below.
% \begin{eqnarray}
% \lefteqn{V_m^{\genie}(\pi_m,\{\textgoth{A}_{k}\}_{k\le m})}\nonumber\\
% &=& R_{m}(\pi_{m},a_m)+\E[V_{m-1}^{\genie}(\pi_{m-1},\{\textgoth{A}_{k}\}_{k\le m-1})|\pi_{m},a_{m}].\nonumber
% \end{eqnarray}
% Note that since the current channel state of both the users ($S_{m}(1)$ and $S_{m}(2)$) are available at the base station at the end of
% the control interval $m$, the belief vector $\pi_{m-1}$ and hence the expected future reward
% $\E[V_{m-1}^{\genie}]$ are independent of the
% scheduling decision $a_{m}$. Therefore, using a proof technique similar to that of
% Proposition~\ref{prop:myoopt}, it can be proved that in any control interval, the net expected reward is maximized by the
% greedy policy in the genie-aided system. This establishes the sum throughput optimality of the greedy policy in the genie-aided
% system as well.
The proposition thus follows.
\end{proof}

{\bf Remarks:} Insights on the result in Proposition~\ref{prop:capacity} can be obtained by examining
the fundamental trade-off when scheduling in the Markov-modeled downlink. In particular, scheduling
must take into account
\begin{itemize}
\item[]\hspace{-12pt}1) data transmission in the current slot, which influences the immediate reward, and
\item[]\hspace{-12pt}2) probing of the channel for future scheduling decisions, which influences the
reward expected in future slots.
\end{itemize}
The optimal schedule strikes a balance between these two objectives (that need not
contradict each other). From the discussion in the proof of Proposition~\ref{prop:capacity}, we see that, in the original
system, when $N=2$, the choice of the user whose channel
is probed becomes irrelevant as far as the optimal future reward is concerned.
%\footnote{As long as one of the users is probed.}
Similarly, in the
genie-aided system, since the channel state information of all the users (general $N$ system) is
sent to the scheduler (with equal delay that is \emph{i.i.d} across the scheduling choice) irrespective of which user was scheduled, the optimal future reward
is independent of the current scheduling decision. This results in the optimality of the greedy
policy in the original and the genie-aided systems and creates a sum capacity equivalence
between these two systems, when $N=2$.

The equivalence with the genie-aided system vanishes when $N>2$, since observing only one user is not enough to
capture an `ON-user', if one exists. This was possible when $N=2$. Thus, when $N>2$, there is room for
throughput improvement when the channel state information of all the users is available at the
scheduler even if there is a delay (the genie-aided system). The genie-aided system sum capacity is
thus an upper bound to the sum capacity of the original system. We record this next.

\begin{corollary}
When $N>2$, the sum capacity, $C_{\summ}(N)$, of the downlink can be bounded as
\begin{eqnarray}
C_{\summ}(2) \le C_{\summ}(N) \le C_{\summ}^{\genie}(N)
%
% \sum_{l=0}^{\infty}\Big[(1-(1-p_s)^N)T^l(p)+(1-p_s)^NT^l(r)\Big]P(D\le l)\prod_{d=0}^{l-1}P(D>d)
\end{eqnarray}
\end{corollary}

\begin{proof}
The lower bound $C_{\summ}(2)$, given in (\ref{eq:csum}), is achieved by the scheduler when, in each slot, it considers only two users
(fixed set) for scheduling and ignores the rest, effectively emulating a two-user downlink.
The upper bound is the sum capacity of the genie-aided system with $N$ users, as given in (\ref{eq:csumgenie}).
% as derived in the proof of Proposition~\ref{prop:capacity}.
\end{proof}

\subsection{Bounds on the Capacity Region of the Downlink}

Define the capacity region of the downlink as the \textit{exhaustive} set of achievable
throughput vectors. Formally, let $\mu_i^{\textgoth{A}}$ denote the throughput of user
$i$ under policy $\textgoth{A}$. Let $I_k(i)$ be the indicator function on whether user $i$ was scheduled
in slot $k$, i.e.,
\begin{eqnarray}
I_k(i)&=&\begin{cases}
1 & \textrm{if~} i=a_k\\
0 & \textrm{otherwise.}
\end{cases}
\end{eqnarray}
Thus
\begin{eqnarray}
\mu_i^\textgoth{A}&=&\lim_{m\rightarrow
\infty}\frac{\E\big[\sum_{k=1}^{m}R^\textgoth{A}_k(\pi_k,a_k)I_k(i)\big]}{m},
\end{eqnarray}
where $R^\textgoth{A}_k(\pi_k,a_k)$ is the immediate reward accrued by the scheduler in slot $k$ under policy
$\textgoth{A}$. The expectation is over the belief vector $\pi_k$ with steady state initial
conditions. Now, the capacity region of the downlink, $\mathcal{C}$, is
defined as the union of the throughput vectors,
$(\mu_1^{\textgoth{A}},\ldots,\mu_{N}^\textgoth{A})$, over all scheduling policies, i.e.,
\begin{eqnarray}
\mathcal{C}&=&\cup_{\textgoth{A}}\{(\mu_1^{\textgoth{A}},\ldots,\mu_{N}^\textgoth{A})\}.
\end{eqnarray}
Let $H_{\convex}(X)$ be the convex
hull of the set of points $X$, defined as
\begin{eqnarray}
\lefteqn{H_{\convex}(X)}\nonumber\\
&=&\Big\{\sum_{i=1}^{n(X)}\beta_{i}x_{i}~\Big|~x_{i}\in X, \beta_{i}\in
\mathbb{R},\beta_{i}\ge 0, \sum_{i=1}^{n(X)}\beta_{i}=1\Big\}.\nonumber
\end{eqnarray}
where $n(X)$ is the cardinality of set $X$. With these definitions we now state our results on the
downlink capacity region.

\begin{proposition}\label{prop:iobounds}
An outer bound on the capacity region of the Markov-modeled downlink with randomly delayed ARQ is given by the complement of the
$N$-dimensional polyhedron $\mathcal{P}$ represented by
\begin{eqnarray}
\mathcal{P}&=&\Big\{(x_1\ge 0, x_2\ge 0\ldots x_N\ge 0):\nonumber\\
&&\sum_{i\in S}x_i \le C_{\summ}^{\genie}(n(S)), \forall S\subseteq
\{1,\ldots N\}\Big\},
\end{eqnarray}
where
\begin{eqnarray}
C_{\summ}^{\genie}(N)\hspace{-8pt}&=&\hspace{-8pt}\sum_{l=0}^{\infty}\Big[(1-(1-p_s)^N)T^l(p)+(1-p_s)^NT^l(r)\Big]\hspace{-3pt}\times\nonumber\\
&&\hspace{10pt}P(D\le l)\prod_{d=0}^{l-1}P(D>d).\nonumber
\end{eqnarray}
%
%
% \begin{eqnarray}
% C(u)&=&\sum_{l=0}^{\infty}\Big[(1-(1-p_s)^u)T^l(p)+(1-p_s)^uT^l(r)\Big]P_L(l)
% \end{eqnarray}
% with
% \begin{eqnarray}
% P_L(l)&=&P(D\le l)\prod_{d=0}^{l-1}P(D>d).
% \end{eqnarray}
An inner bound on the capacity region is given by the set of points $(x_1,\ldots,x_N)$ such that
\begin{eqnarray}
\lefteqn{(x_1,\ldots,x_N)}\nonumber\\
&\in&H_{\convex}(O,\{X_i\}_{\forall i\in \{1,\ldots, N\}}, \{Y_{j,k}\}_{\forall j,k\in \{1,\ldots,
N\}, j\neq k})\nonumber\\
\end{eqnarray}
where $O,X_i,Y_{j,k}\in\mathbb{R}^N$. $O$ is the origin $(0,\ldots, 0)$. $X_i=(0,\ldots,0,p_s,0,\ldots,0)$ with $p_s$ at
the $i^{th}$ location. $Y_{j,k,j\neq k}=(0,\ldots,0,\frac{C_{\summ}(2)}{2},0,\ldots,0,\frac{C_{\summ}(2)}{2},0,\ldots,0)$
with $\frac{C_{\summ}(2)}{2}$ at locations $j$ and $k$, where
\begin{eqnarray}
C_{\summ}(2)&=&\sum_{l=0}^{\infty}\Big[p_{s}T^l(p)+(1-p_s)p_s\Big]\times\nonumber\\
&&\hspace{17pt}P(D\le l)\prod_{d=0}^{l-1}P(D>d).\nonumber
\end{eqnarray}
% \begin{eqnarray}
% \mathcal{L}&=&\Big\{(x_1\ge 0, x_2\ge 0\ldots x_N\ge 0):\sum_{i\in S}x_i \le p_s, \forall S\subseteq
% \{1,\ldots N\}\Big\}.
% \end{eqnarray}
\end{proposition}

\begin{figure}
\centering
\psfrag{2}[cc]{\small{$ N=2 $}}
\psfrag{3}[cc]{\small{$ N=3 $}}
\psfrag{x1}[ll]{\tiny{$x_1$}}
\psfrag{x2}[ll]{\tiny{$x_2$}}
\psfrag{x3}[ll]{\tiny{$x_3$}}
\psfrag{o}[rr]{\tiny{$(0,0)$}}
\psfrag{p}[ll]{\tiny{$(\frac{C_{\summ}(2)}{2},\frac{C_{\summ}(2)}{2})$}}
\psfrag{p1}[ll]{\tiny{$(p_s,0)$}}
\psfrag{p2}[rr]{\tiny{$(0,p_s)$}}
\psfrag{O}[ll]{\tiny{$(0,0,0)$}}
\psfrag{P1}[ll]{\tiny{$(p_s,0,0)$}}
\psfrag{P2}[ll]{\tiny{$(0,p_s,0)$}}
\psfrag{P3}[rr]{\tiny{$(0,0,p_s)$}}
\psfrag{C1}[ll]{\tiny{$x_2=C_{\summ}^{\genie}(1)=p_s$ }}
\psfrag{C2}[ll]{\tiny{$x_1+x_2=C_{\summ}^{\genie}(2)$ }}
%\psfrag{c2}[ll]{\tiny{$x_2+x_3=C_{\summ}^{\genie}(2)$}}
\psfrag{C3}[ll]{\tiny{$x_1+x_2+x_3=C_{\summ}^{\genie}(3)$ }}
\psfrag{y1}[ll]{\tiny{$(\frac{C_{\summ}(2)}{2},\frac{C_{\summ}(2)}{2},0)$}}
\psfrag{y2}[rr]{\tiny{$(0,\frac{C_{\summ}(2)}{2},\frac{C_{\summ}(2)}{2})$}}
\psfrag{y3}[cc]{\tiny{$(\frac{C_{\summ}(2)}{2},0,\frac{C_{\summ}(2)}{2})$}}
\psfrag{ob}[ll]{\scriptsize{Outer bound}}
\psfrag{ib}[ll]{\scriptsize{Inner bound}}
\includegraphics[width=.35\textwidth]{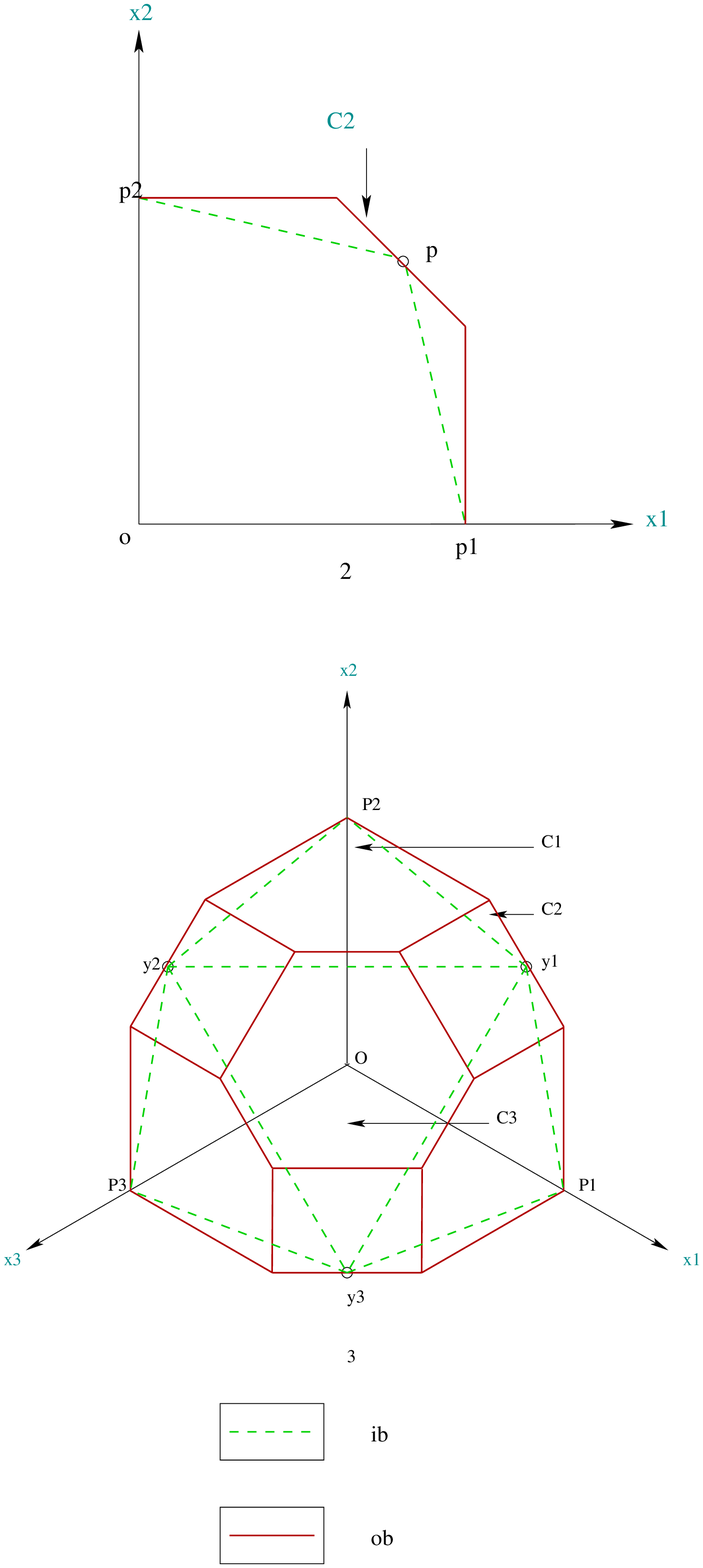}
\caption{Illustration of bounds on the capacity region of the downlink with randomly delayed ARQ when $N=2$ and when $N=3$.}\label{fig:Fig3}
\end{figure}

\begin{proof}
Considering the genie-aided system, for any policy $\textgoth{A}$, let the throughput vector
be denoted by $(\mu_{1}^{\textgoth{A},\genie},\ldots,\mu_{N}^{\textgoth{A},\genie})$. For a subset
of users $S\subseteq\{1\ldots N\}$, by the definition of sum capacity, we have
\begin{eqnarray}
\sum_{i\in S}\mu_{i}^{\textgoth{A},\genie}&\le&C_{\summ}^{\genie}(n(S)).
\end{eqnarray}
This establishes the complement of the polyhedron $\mathcal{P}$ as an outer bound on the capacity
region of the genie-aided system, and by extension, an outer bound on the capacity region of the
original system.

Now, consider the inner bound $H_{\convex}(O,\{X_i\}_{\forall i\in \{1,\ldots, N\}},
\{Y_{j,k}\}_{\forall j,k\in \{1,\ldots, N\}, j\neq k})$. In the original system, throughput vector
$X_i=(0,\ldots,0,p_s,0,\ldots,0)$ can be achieved by scheduling to user $i$ at all
times. Recall that the greedy policy achieves the sum capacity when $N=2$. Also the sum throughput $C_{\summ}(2)$ is split equally between the
two users thanks to the inherent symmetry between users. Thus throughput vector
$Y_{j,k,j\neq k}=(0,\ldots,0,\frac{C_{\summ}(2)}{2},0,\ldots,0,\frac{C_{\summ}(2)}{2},0,\ldots,0)$
can be achieved by greedy scheduling over the users $j$ and $k$ alone at all slots. Throughput vector $O$
corresponds to idling in every slot. Therefore,
any throughput vector in the convex hull $H_{\convex}(O,\{X_i\}_{\forall i\in \{1,\ldots, N\}},
\{Y_{j,k}\}_{\forall j,k\in \{1,\ldots, N\}, j\neq k})$ can be achieved by time sharing between the
policies that achieve throughput vectors $\in \{O,X_{i},Y_{j,k,j\neq k}\}$. This establishes the result on the inner bound.
\end{proof}
\figref{Fig3} illustrates the capacity region bounds from Proposition~\ref{prop:iobounds} when $N=2$ and when $N=3$.

For the special case of $N=2$ users and deterministic ARQ feedback delay, $D=d$, we obtain the
exact capacity region of the genie-aided system and hence tighter bounds
to the capacity region of the original system.

\begin{proposition}\label{prop:convregion}
For $N=2$ users, with a deterministic ARQ delay of $D=d,~d\ge 0$ slots, the capacity
region of the genie-aided system is given by the set of points
$(x_{1},x_{2})$ such that
\begin{eqnarray}\label{eq:conv}
(x_{1},x_{2}) \hspace{-5pt}&\in& \hspace{-5pt} H_{\convex}(O, X_1, Z_1, Z_2, X_2)\nonumber\\
\textrm{where  } O \hspace{-5pt}&=& \hspace{-5pt}(0,0)\nonumber\\
X_1 \hspace{-5pt}&=& \hspace{-5pt}(p_s,0)\nonumber\\
X_2 \hspace{-5pt}&=& \hspace{-5pt}(0,p_s)\nonumber\\
Z_1 \hspace{-5pt}&=& \hspace{-5pt}\big(p_{s}T^d(p)+(1-p_s)^{2}T^d(r), (1-p_s)p_{s}T^d(p)\big)\nonumber\\
Z_2 \hspace{-5pt}&=& \hspace{-5pt}\big((1-p_s)p_{s}T^d(p), p_{s}T^d(p)+(1-p_s)^{2}T^d(r)\big).\nonumber\\
\end{eqnarray}
% and an inner bound is given by the set of points $(\lambda_{1},\lambda_{2})$ such that
% \begin{eqnarray}\label{eq:achiev}
% (\lambda_{1},\lambda_{2}) &\in& H_{\convex}(O, A_1, E, A_2)\nonumber\\
% \textrm{where  }E&=&(\frac{C_{\summ}}{2},\frac{C_{\summ}}{2})\nonumber
% \end{eqnarray}
% with $C_{\summ}=p_{s}T^d(p)+(1-p_s)p_s$, the sum capacity of the system.
\end{proposition}

\begin{proof}
The relative positions of the points $X_1,~ X_2,~Z_1,~Z_2$ and $O$ are illustrated in
\figref{Fig5}.
% %\putFrag{Fig5}{Relative locations of points $A_1,~ A_2,~Z_1,~Z_2, ~E$ and $O$}{9}{}
% \putFrag{bounds2d}{Inner and outer bounds under random ARQ delay when $N=2$ users}{10}{}
% %\putFrag{bounds3d}{Inner and outer bounds under random ARQ delay when $N=3$ users}{12}{}
% \putFrag{bounds2d_new}{Relative locations of points $A_1,~ A_2,~Z_1,~Z_2, ~S$ and $O$}{10}{}

We proceed by first showing that the region complementary to $H_{\convex}(O, X_1, Z_1, Z_2, X_2)$ is
an outer bound on the capacity
region of the genie-aided downlink. Consider a
broad class of schedulers in the genie-aided system, with each member identified by the parameters
$\alpha_{i}\in[0,1],~i\in\{1,\ldots, 4\}$. A member of this class obeys the following decision
logic at slot $t$:
\vspace{5pt}
\begin{itemize}
\item If $\begin{bmatrix}
S_{t+d+1}(1)\\
S_{t+d+1}(2)\end{bmatrix}$=$\begin{bmatrix}
0\\
0\end{bmatrix}$, then schedule user 1 with probability $\alpha_{1}$ and user 2 w.p. $1-\alpha_{1}$.
\vspace{5pt}
\item If $\begin{bmatrix}
S_{t+d+1}(1)\\
S_{t+d+1}(2)\end{bmatrix}$=$\begin{bmatrix}
0\\
1\end{bmatrix}$, then
$a_{t}=\left\{\begin{array}{ll}
1~ w.p.~ \alpha_2\\
2~ w.p.~ 1-\alpha_2
\end{array}\right.$
\vspace{5pt}

\item If $\begin{bmatrix}
S_{t+d+1}(1)\\
S_{t+d+1}(2)\end{bmatrix}$=$\begin{bmatrix}
1\\
0\end{bmatrix}$, then
$a_{t}=\left\{\begin{array}{ll}
1~ w.p.~ \alpha_3\\
2~ w.p.~ 1-\alpha_3
\end{array}\right.$
\vspace{5pt}

\item If $\begin{bmatrix}
S_{t+d+1}(1)\\
S_{t+d+1}(2)\end{bmatrix}$=$\begin{bmatrix}
1\\
1\end{bmatrix}$, then
$a_{t}=\left\{\begin{array}{ll}
1~ w.p.~ \alpha_4\\
2~ w.p.~ 1-\alpha_4
\end{array}\right.$

\end{itemize}
\vspace{5pt}
% Since $\begin{bmatrix}
% S_{k+1}(1)\\
% S_{k+1}(2)\end{bmatrix}$ is a sufficient statistic for $\begin{bmatrix}
% S_{k}(1)\\
% S_{k}(2)\end{bmatrix}$,
Note that, thanks to the first order Markovian nature of the underlying channels, any scheduling
policy in the genie-aided system falls under the above class
of schedulers or will have a member of this class achieving the same throughput vector as itself.
We now proceed to show that the throughput vector achieved by any member of this class
belongs to $H_{\convex}(O, X_1, Z_1, Z_2, X_2)$.

\begin{figure}
\centering
\psfrag{x1}[ll]{\tiny{$x_1$}}
\psfrag{x2}[ll]{\tiny{$x_2$}}
\psfrag{o}[rr]{\tiny{$O=(0,0)$}}
\psfrag{p}[ll]{\tiny{$Y_{1,2}=(\frac{C_{\summ}(2)}{2},\frac{C_{\summ}(2)}{2})$}}
\psfrag{pp}[ll]{\scriptsize{Outer bound from Proposition~\ref{prop:iobounds}}}
\psfrag{z1}[ll]{\tiny{$Z_1$}}
\psfrag{z2}[ll]{\tiny{$Z_2$}}
\psfrag{p1}[ll]{\tiny{$X_1=(p_s,0)$}}
\psfrag{p2}[rr]{\tiny{$X_2=(0,p_s)$}}
\psfrag{v1}[ll]{\tiny{$(p_s,C_{\summ}-p_s)$}}
\psfrag{v2}[cc]{\tiny{$(C_{\summ}-p_s,p_s)$}}
\psfrag{o1}[ll]{\scriptsize{Capacity region of the genie-aided system}}
\psfrag{o2}[ll]{\scriptsize{and a tighter outer bound on the capacity}}
\psfrag{o3}[ll]{\scriptsize{region of the original system}}
% \psfrag{ob}[ll]{$\begin{array}{c}\vspace{-5pt}\textrm{\small{capacity region of the genie-aided
% system}}\\\vspace{-5pt}\textrm{\small{and a tighter outer bound to the
% capacity}}\\\vspace{-5pt}\textrm{\small{region of the original system}}\end{array}$}
% \psfrag{ob}[ll]{$\begin{array}{c}\vspace{-5pt}\textrm{\small{Outer bound to  remains on
% top.}}\\\vspace{-5pt}\textrm{\tiny{It will be rescheduled}}\\
% \vspace{-5pt}\textrm{\tiny{in the next slot}}\end{array}$}
%
\psfrag{ib}[ll]{\scriptsize{Inner bound}}

%
% \psfrag{O}[ll]{\tiny{$(0,0,0)$}}
% \psfrag{P1}[ll]{\tiny{$(p_s,0,0)$}}
% \psfrag{P2}[ll]{\tiny{$(0,p_s,0)$}}
% \psfrag{P3}[rr]{\tiny{$(0,0,p_s)$}}
% \psfrag{C1}[ll]{\tiny{$x_2=C_{\summ}^{\genie}(1)=p_s$}}
% \psfrag{C2}[ll]{\tiny{$x_1+x_2=C_{\summ}^{\genie}(2)$}}
% %\psfrag{c2}[ll]{\tiny{$x_2+x_3=C_{\summ}^{\genie}(2)$}}
% \psfrag{C3}[ll]{\tiny{$x_1+x_2+x_3=C_{\summ}^{\genie}(3)$}}
% \psfrag{y1}[ll]{\tiny{$(\frac{C_{\summ}(2)}{2},\frac{C_{\summ}(2)}{2},0)$}}
% \psfrag{y2}[rr]{\tiny{$(0,\frac{C_{\summ}(2)}{2},\frac{C_{\summ}(2)}{2})$}}
% \psfrag{y3}[cc]{\tiny{$(\frac{C_{\summ}(2)}{2},0,\frac{C_{\summ}(2)}{2})$}}
\includegraphics[width=.275\textwidth]{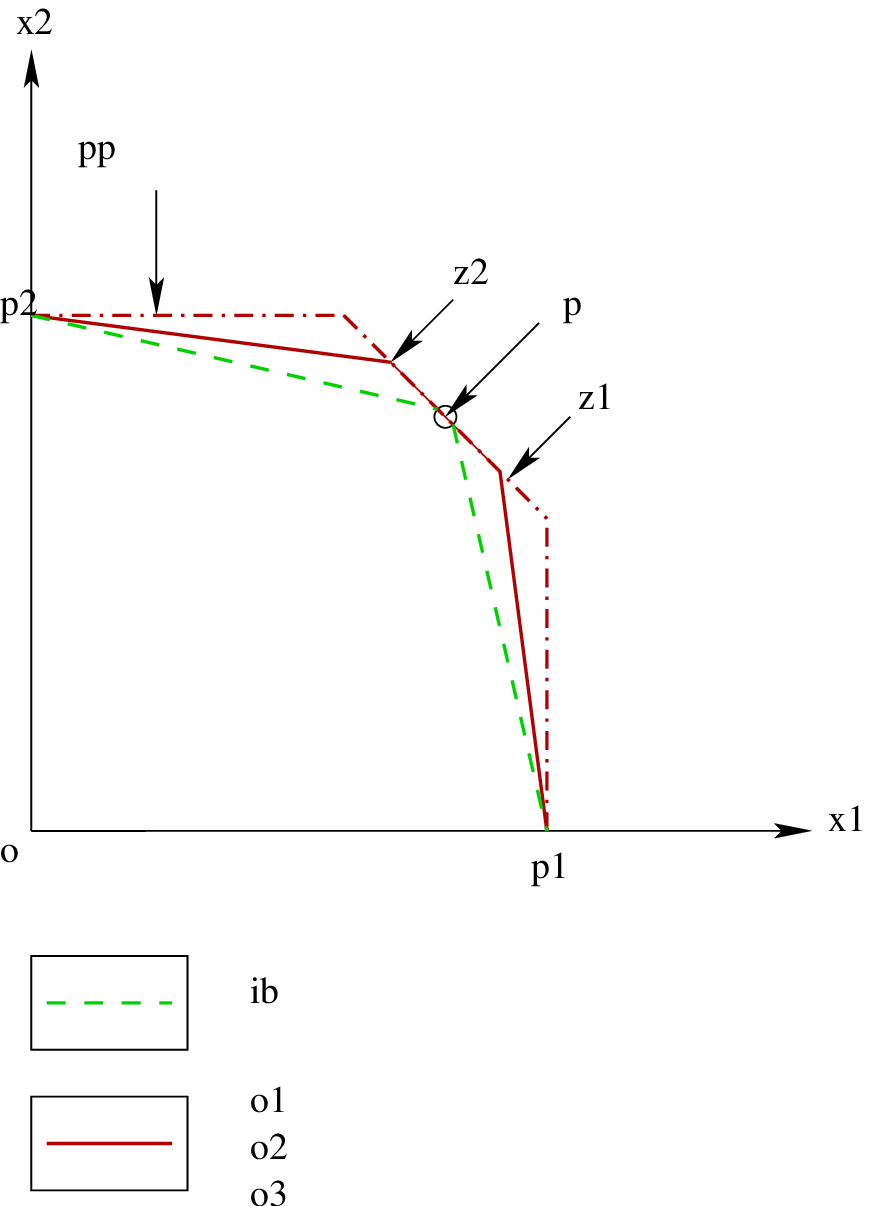}
\caption{Illustration of the capacity region of the genie-aided system and tighter bounds on the
capacity region of the original system when $N=2$, with deterministic ARQ delay.}\label{fig:Fig5}
\end{figure}

With $\vec{\alpha}=\{\alpha_{1},\ldots, \alpha_{4}\}\in[0,1]^4$ fixed, the throughput for user 1 is given by
\begin{eqnarray}\label{eq:mu1}
\lefteqn{\mu^{\vec{\alpha},\genie}_{1}}\nonumber\\
&=&\hspace{-12pt}\sum_{i,j\in\{0,1\}}P\Big(\begin{bmatrix}
S_{t+d+1}(1)\\
S_{t+d+1}(2)\end{bmatrix}=\begin{bmatrix}
i\\
j\end{bmatrix}\Big)\times\nonumber\\
&&\hspace{-19pt}P\Big(a_{t}=1\hspace{-1pt}\Big|\hspace{-3pt}\begin{bmatrix}
S_{t+d+1}(1)\\
S_{t+d+1}(2)\end{bmatrix}\hspace{-3pt}=\hspace{-3pt}\begin{bmatrix}
i\\
j\end{bmatrix}\Big)P(S_{t}(1)=1|S_{t+d+1}(1)=i)\nonumber\\
&=&\hspace{-5pt}(1-p_s)^{2}\alpha_{1}T^d(r)+(1-p_s)p_{s}\alpha_{2}T^d(r)\nonumber\\
&&\hspace{-5pt}+\hspace{3pt}p_s(1-p_s)\alpha_{3}T^d(p)+p_{s}^{2}\alpha_{4}T^d(p),
\end{eqnarray}
with $p_s=\frac{r}{1-(p-r)}$. Similarly,
\begin{eqnarray}\label{eq:mu2}
\lefteqn{\mu^{\vec{\alpha},\genie}_{2}}\nonumber\\
&=&(1-p_s)^{2}(1-\alpha_{1})T^d(r)+(1-p_s)p_{s}(1-\alpha_{2})T^d(p)\nonumber\\
&&+~p_s(1-p_s)(1-\alpha_{3})T^d(r)+p_{s}^{2}(1-\alpha_{4})T^d(p).\nonumber\\
\end{eqnarray}
For notational simplicity, we will henceforth denote the throughputs simply by $\mu_1$ and $\mu_2$. The sum
throughput is now given by
\begin{eqnarray}
\mu_{1}+\mu_{2}=p_s+(1-p_s)p_{s}(T^d(p)-T^d(r))(\alpha_3-\alpha_2).
%\mu_1+\mu_2=\frac{(1-p)^{2}r+r^{2}p+(1-p)rp(1+\alpha_3-\alpha_2)+(1-p)r^{2}(1-(\alpha_3-\alpha_2))}{(1-(p-r))^2}
\end{eqnarray}

Note that the values of $\alpha_{1}$ and $\alpha_{4}$
are irrelevant from the sum throughput point of view. Consider the following two cases.\\
\textit{Case 1, when $\alpha_3\le\alpha_2$:}  
\begin{eqnarray}
0\le\mu_1+\mu_2\le p_s.\nonumber
\end{eqnarray}
Since $X_1(1)+X_1(2)=X_2(1)+X_2(2)=p_s$, we have 
\begin{eqnarray}\label{eq:hull1}
(\mu_1,\mu_2)\in H_{\convex}(O,X_1,X_2).
\end{eqnarray}
\textit{Case 2, when $\alpha_3>\alpha_2$:}
\begin{eqnarray}
p_s<\mu_1+\mu_2&\le& p_s+(1-p_s)p_{s}(T^d(p)-T^d(r))\nonumber\\
&=&p_{s}T^d(p)+(1-p_s)p_s.\nonumber
\end{eqnarray}
Since $Z_1(1)+Z_1(2)=Z_2(1)+Z_2(2)=p_{s}T^d(p)+(1-p_s)^{2}T^d(r)+(1-p_s)p_{s}T^d(p)=p_{s}T^d(p)+(1-p_s)p_s$, we can find points
$E_{X_{1}Z_{1}}$ and $E_{X_{2}Z_{2}}$ on edges $X_{1}Z_{1}$ and $X_{2}Z_{2}$, respectively, such
that $E_{X_{1}Z_{1}}(1)+E_{X_{1}Z_{1}}(2)=E_{X_{2}Z_{2}}(1)+E_{X_{2}Z_{2}}(2)=\mu_1+\mu_2$. Any
point $P_{X_{1}Z_{1}}$ on the edge $X_{1}Z_{1}$ can be written as a convex combination of points $X_1$ and $Z_1$,
i.e., $\exists~ \beta\in[0,1]$ such that
\begin{eqnarray}
P_{X_{1}Z_{1}}&=&X_{1}\beta+Z_{1}(1-\beta)\nonumber\\
&=&\Big(p_{s}\beta+ (p_{s}T^d(p)+(1-p_s)^{2}T^d(r))(1-\beta),\nonumber\\
&&\hspace{10pt}(1-p_s)p_{s}T^d(p)(1-\beta)\Big).\nonumber
\end{eqnarray}
With $\beta=1-(\alpha_3-\alpha_2)$, we have $P_{X_{1}Z_{1}}(1)+P_{X_{1}Z_{1}}(2)=\mu_1+\mu_2$. Thus 
\begin{eqnarray}
E_{X_{1}Z_{1}}\hspace{-10pt}&=&\hspace{-10pt}\Big(p_{s}(1-(\alpha_3-\alpha_2))+(p_{s}T^d(p)+(1-p_s)^{2}T^d(r))\times\nonumber\\
&&(\alpha_3-\alpha_2), (1-p_s)p_{s}T^d(p)(\alpha_3-\alpha_2)\Big).\nonumber
\end{eqnarray}
Due to the symmetry between $X_{1}$, $Z_{1}$ and $X_{2}$, $Z_{2}$, we have
$E_{X_{2}Z_{2}}=(E_{X_{1}Z_{1}}(2),~E_{X_{1}Z_{1}}(1))$. Using $\mu_{1}$ from~(\ref{eq:mu1}), it can
be shown that, for any $\alpha_{i\in\{1\ldots 4\}}\in[0,1]$ with $\alpha_3>\alpha_2$,
\begin{eqnarray}\label{eq:convex}
E_{X_{2}Z_{2}}(1)\le\mu_1\le E_{X_{1}Z_{1}}(1).
\end{eqnarray}
Since $E_{X_{1}Z_{1}}(1)+E_{X_{1}Z_{1}}(2)=E_{X_{2}Z_{2}}(1)+E_{X_{2}Z_{2}}(2)=\mu_1+\mu_2$,
(\ref{eq:convex}) translates to
\begin{eqnarray}
(\mu_1,\mu_2)\in H_{\convex}(E_{X_{1}Z_{1}},E_{X_{2}Z_{2}}).\nonumber
\end{eqnarray}
The above relation, along with the fact that $E_{X_{1}Z_{1}}\in H_{\convex}(X_1,Z_1)$ and
$E_{X_{2}Z_{2}}\in H_{\convex}(X_2,Z_2)$, yields
\begin{eqnarray}\label{eq:hull2}
(\mu_1,\mu_2)\in H_{\convex}(X_1,Z_1,Z_2,X_2).
\end{eqnarray}
Combining the results in~(\ref{eq:hull1}) and~(\ref{eq:hull2}), we establish that the region
complementary to $H_{\convex}(O, X_1, Z_1, Z_2, X_2)$ is an outer bound on the capacity region of
the genie-aided system. 

Revisiting the class of schedulers identified by $\vec{\alpha}$, it
can be shown from (\ref{eq:mu1}) and (\ref{eq:mu2}) that a
scheduler with $\vec{\alpha}=\{1,0,1,1\}$ achieves a throughput vector
$(\mu_1,\mu_2)=Z_1=\big(p_{s}T^d(p)+(1-p_s)^{2}T^d(r), (1-p_s)p_{s}T^d(p)\big)$. Similarly, a
scheduler with $\vec{\alpha}=\{0,0,1,0\}$ achieves a throughput vector
$(\mu_1,\mu_2)=Z_2=\big((1-p_s)p_{s}T^d(p), p_{s}T^d(p)+(1-p_s)^{2}T^d(r)\big)$. Throughput vectors $X_1$
or $X_2$ can be achieved by scheduling to only user 1 or 2, respectively, at all times. Thus any throughput 
vector within the region $H_{\convex}(O, X_1, Z_1, Z_2, X_2)$ can be supported by time sharing between
the schedulers that achieve throughput vector $\in\{O,X_1,Z_1,Z_2,X_2\}$. This establishes $H_{\convex}(O, X_1, Z_1,
Z_2, X_2)$ as an inner bound on the capacity region of the genie-aided system.

Combining the outer and inner bound results establishes the proposition.
%
%
% Since the sum capacity is achieved by the greedy policy for $N=2$ users, and since the users in the
% system are inherently symmetric, the rate vector $E$ can be achieved by the greedy policy. Rate vectors $X_1$ and $X_2$ can
% be achieved by scheduling transmission to only user 1 or 2, respectively, in every slot.
% Any rate on the edges $X_{1}S$ or $X_{2}S$ can be achieved by time sharing between the
% corresponding two policies. From the definition of the convex hull, for any arrival rate vector
% belonging to the region $H_{\convex}(O, X_1, E, X_2)$, we can always find a stabilizing rate vector on one of the edges $X_{1}S$
% or $X_{2}S$. The proposition thus follows.
\end{proof}

We now report tighter bounds on the capacity region of the original system, when
$N=2$ and the ARQ delay is deterministic.
\begin{corollary}\label{cor:boundcor}
For $N=2$ users, with a deterministic ARQ delay of $D=d,~d\ge 0$ slots, an outer bound on the capacity
region of the original system is given by the set of points
$(x_{1},x_{2})$ such that
\begin{eqnarray}\label{eq:conv}
(x_{1},x_{2}) &\notin& H_{\convex}(O, X_1, Z_1, Z_2, X_2)
\nonumber\\
\textrm{where  } O&=&(0,0)\nonumber\\
X_1&=&(p_s,0)\nonumber\\
X_2&=&(0,p_s)\nonumber\\
Z_1&=&\big(p_{s}T^d(p)+(1-p_s)^{2}T^d(r), (1-p_s)p_{s}T^d(p)\big)\nonumber\\
Z_2&=&\big((1-p_s)p_{s}T^d(p), p_{s}T^d(p)+(1-p_s)^{2}T^d(r)\big)\nonumber\\
\end{eqnarray}
and an inner bound is given by the set of points $(x_{1},x_{2})$ such that
\begin{eqnarray}\label{eq:achiev}
(x_{1},x_{2}) &\in& H_{\convex}(O, X_1, Y_{1,2}, X_2)\nonumber\\
\textrm{where  }Y_{1,2}&=&(\frac{C_{\summ}(2)}{2},\frac{C_{\summ}(2)}{2})\nonumber
\end{eqnarray}
with $C_{\summ}(2)=p_{s}T^d(p)+(1-p_s)p_s$, the sum capacity of the system.
\end{corollary}

\begin{proof}
The outer bound is the region complementary to the capacity region of the genie-aided system
reported in Proposition~\ref{prop:convregion}. The inner bound was obtained in
Proposition~\ref{prop:iobounds} with $C_{\summ}(2)$ from (\ref{eq:csum}) re-derived using $P(D=d)=1$.
\end{proof}
\figref{Fig5} illustrates the improved outer bound from Corollary~\ref{cor:boundcor}
along with the bounds derived in Proposition~\ref{prop:iobounds}.

\section{Conclusion}\label{sec:conclusion}
We addressed the problem of opportunistic multiuser scheduling for a system consisting of a base station or access point transmitting to users within its domain. We model the downlink channels by two-state Markov chains, with ON and OFF states, and assume that the data destined for each user is infinitely backlogged. We allow for the ARQ feedback from each user to the base station to be randomly (\textit{i.i.d.} over all users) delayed. For the case of two users in the system, we showed that the greedy policy is sum throughput optimal for any distribution of the ARQ feedback delay. However, for more than two users, there exists scenarios for which the greedy policy is not optimal. Nevertheless, extensive numerical experiments suggest that the greedy policy has near optimal performance. Encouraged by this, we studied the structure of the
greedy policy and showed that it can be implemented by a simple algorithm that does not
require the statistics of the underlying Markov channel nor the ARQ feedback delay, thus making it
robust against errors in estimation of these statistics. Focusing on the fundamental limits of the
downlink system, we obtained an elegant closed form expression for the sum
capacity of the two-user downlink and derived inner and outer bounds
on the capacity region of the Markov-modeled downlink with randomly delayed ARQ feedback.

In summary, we addressed opportunistic multiuser scheduling based on existing ARQ feedback
mechanisms, while taking into account an important non-ideality in the feedback channel - the random delay. We
studied this scheduling problem by examining various aspects of the `easy to implement'
greedy policy and by establishing fundamental limits on the downlink system performance. We believe that the work we have initiated here, along with the proof techniques we have developed, could be the first steps towards studying the joint channel learning - scheduling problem under more general scenarios: such as, when the users have heterogeneous demands, when the queues are non-backlogged with random packet arrivals.

\appendices

\section{Proof of Lemma~\ref{le:Tprop}}\label{app:lemma}

Recall the definition of the $u-$step belief evolution operator:
$T^u(x)=T(T^{(u-1)}(x))=T^{(u-1)}(T(x))$ with $T(x)=xp+(1-x)r=x(p-r)+r$ and $T^{0}(x)=x$ for $x\in[0,1]$ and $u\in\{0,1,2,\ldots\}$.
For $u\in\{1,2,\ldots\}$, $x \in [0,1]$,
\begin{eqnarray}
T^{u}(p)&=&T^{(u-1)}(p)p+(1-T^{(u-1)}(p))r\nonumber\\
T^{(u+1)}(x)&=&T^u(x)p+(1-T^u(x))r\nonumber\\
T^u(p)-T^{(u+1)}(x)&=&(p-r)(T^{(u-1)}(p)-T^{u}(x)).
\end{eqnarray}
Thus if, for $u\in\{1,2,\ldots\}$, $T^{(u-1)}(p)-T^{u}(x)\ge 0$, then, since $p>r$, we have
$T^u(p)-T^{(u+1)}(x)\ge 0$. By induction, using $p\ge T(x)=xp+(1-x)r$ for any $x\in [0,1]$, we have
$T^{u}(p)\ge T^{u+1}(x)$ for any $u\in\{0,1,2,\ldots\}$ and $x\in [0,1]$. The second inequality in
the lemma can be proved along the same lines using $r\le T(x)=xp+(1-x)r$. 

Consider the third inequality. By definition, for any $x,y\in [0,1]$,
$T^u(x)-T^u(y)=(p-r)(T^{(u-1)}(x)-T^{(u-1)}(y))$. Thus, if $T^{(u-1)}(x)-T^{(u-1)}(y)$,
then $T^u(x)-T^u(y)\ge 0$. When $x\ge y$, by induction, $T^u(x)-T^u(y)\ge 0$ for any
$u\in\{0,1,2,\ldots\}$. This establishes the third inequality.

Considering the last inequality, the belief evolution operator can be expressed as
\begin{eqnarray}
T^u(x)&=&T(T^{(u-1)}(x))=T(T(T^{(u-2)}(x)))\nonumber\\
&=&x(p-r)^u+r(\frac{1-(p-r)^u}{1-(p-r)})\nonumber\\
&=&\frac{r}{1-(p-r)}+(p-r)^u\big(x-\frac{r}{1-(p-r)}\big)\nonumber\\
\end{eqnarray}
for $u\in\{0,1,2,\ldots\}$ and $x\in[0,1]$.
Thus $T^u(p)=\frac{r}{1-(p-r)}+(p-r)^u\big[\frac{(p-r)(1-p)}{1-(p-r)}\big]$. Note that, since $p>r$,
$T^u(p)\ge \frac{r}{1-(p-r)}$. Also,
$T^u(r)=\frac{r}{1-(p-r)}-(p-r)^u\big[\frac{(p-r)r}{1-(p-r)}\big]\le \frac{r}{1-(p-r)}$. This
establishes the last inequality in the lemma.

\section{Proof of Corollary~\ref{necorr}}\label{app:cor2pt5}

The proof proceeds closely follows that of Proposition~\ref{prop:myoopt}. Recall the quantities $f_k,F_k,\tau_k,k_1,k_2,l_1,l_2$ from the proof of Proposition~\ref{prop:myoopt}. Consider a slot $t<m$ with the sequence of past actions given by 
$\vec{a}_{t+1}=\{a_m\ldots a_{t+1}\}$. The proof proceeds in two steps. In step 1, we show that the greedy decision in slot
$t$, given the ARQ feedback and the scheduling decision from slot $\min(k_1,k_2)$, is independent of the
feedback and scheduling decision corresponding to slot $\max(k_1,k_2)$. 
In step $2$, we show that, if the greedy policy is implemented in slot $t$, then the expected immediate reward in slot $t$ is
independent of the scheduling decisions $\vec{a}_{t+1}$. We then provide induction based arguments to
establish the proposition. 

\textit{Step 1:} Let $\vec{F}_{t+1}:=\{F_m,F_{m-1},\ldots,F_{t+1}\}$ and
$\vec{\tau}_{t+1}:=\{\tau_{m},\tau_{m-1},\ldots,\tau_{t+1}\}$. The greedy decision in slot $t$, conditioned on
the past feedback and scheduling decisions is given by
\begin{eqnarray}
\hat{a}_{t}|_{\vec{F}_{t+1},\vec{\tau}_{t+1},\vec{a}_{t+1},\pi_m}&=&\hat{a}_{t}|_{f_{k_1},f_{k_2},l_1,l_2,\vec{a}_{t+1},\pi_m}.
\end{eqnarray}
The preceding equation comes directly from the first order Markovian property of the underlying
channels.
% ollowing argument: $\hat{a}_t$ is a function of $\pi_t$. $\pi_t(1)$ (likewise,
% $\pi_t(2)$) is independent of $\{_{k\neq k_1},\tau_{k\neq k_1^*}\}$ (likewise, $\{f_{k\neq
% k_2^*},\tau_{k\neq k_2^*}\}$) given $\{f_{k_1},\tau_{k_1^*},\vec{a}_{t+1}\}$
% (likewise, $\{f_{k_2^*},\tau_{k_2^*},\vec{a}_{t+1}\}$) due to the first order Markovian property of the downlink
% channel. 
Consider the case when $k_1<k_2\le m$ ($\Rightarrow l_1<l_2$) or $k_1=k_2= \emptyset$ ($\Rightarrow
l_1=l_2=\emptyset$). The belief values in slot $t$ as a function of feedback $f_{k_1}$ and
$f_{k_2}$ is given below:
\begin{eqnarray}\label{eq:pi22}
\lefteqn{(\pi_t(1),\pi_t(2))}\nonumber\\
&=&\hspace{-10pt}\begin{cases}
\hspace{-2pt}(T^{{l_1}}(p),T^{{l_2}}(p)), & \hspace{-7pt} \mbox{if } f_{k_1}=1,f_{k_2}=1 \\
\hspace{-2pt}(T^{{l_1}}(p),T^{{l_2}}(r)), & \hspace{-7pt} \mbox{if } f_{k_1}=1,f_{k_2}=0 \\
\hspace{-2pt}(T^{{l_1}}(p),T^{(m-t)}(\pi_m(2))), & \hspace{-7pt} \mbox{if } f_{k_1}=1,k_2=\emptyset\\
\hspace{-2pt}(T^{{l_1}}(r),T^{{l_2}}(p)), & \hspace{-7pt} \mbox{if } f_{k_1}=0,f_{k_2}=1\\
\hspace{-2pt}(T^{{l_1}}(r),T^{{l_2}}(r)), & \hspace{-7pt} \mbox{if } f_{k_1}=0,f_{k_2}=0\\
\hspace{-2pt}(T^{{l_1}}(r),T^{(m-t)}(\pi_m(2))), & \hspace{-7pt} \mbox{if } f_{k_1}=0,k_2=\emptyset\\
\hspace{-2pt}(T^{(m-t)}(\pi_m(1)),T^{(m-t)}(\pi_m(2))), & \hspace{-7pt} \mbox{if } k_1=\emptyset,k_2=\emptyset\\
\end{cases}\nonumber\\
\end{eqnarray}
Now, from the definition of $T^k(.)$ and using the fact that $p<r$, the following ineualities can be readily verified.
For $u\in\{1,3,5\ldots\}$, $v>u$ and $x\in[0,1]$,
\begin{eqnarray}
T^u(p)&\ge& T^v(x)\nonumber\\
T^u(r)&\le& T^v(x).
\end{eqnarray}
For $u\in\{0,2,4\ldots\}$, $v>u$ and $x\in[0,1]$,
\begin{eqnarray}
T^u(p)&\le& T^v(x)\nonumber\\
T^u(r)&\ge& T^v(x).
\end{eqnarray}
The preceding inequalities essentially result from the oscillating nature of the evolution of belief values in a negatively correlated Markov channel. 
Using these inequalities and (\ref{eq:pi22}), the greedy decision in slot $t$ can be written as
\begin{eqnarray}\label{eq:ahat_2}
\lefteqn{\hat{a}_{t}|_{f_{k_1},f_{k_2},l_1,l_2,\vec{a}_{t+1},\pi_m}}\nonumber\\
&=&\begin{cases}
\begin{cases}
1, & \mbox{if }f_{k_1}=1\\
2, & \mbox{if }f_{k_1}=0\\
\end{cases}, \mbox{if }k_1\neq\emptyset, l_1\mbox{ is odd}\\
\begin{cases}
2, & \mbox{if }f_{k_1}=1\\
1, & \mbox{if }f_{k_1}=0\\
\end{cases}, \mbox{if }k_1\neq\emptyset, l_1\mbox{ is even}\\
\begin{cases}
\arg\max_{i\in\{1,2\}}(\pi_m(i)), & \hspace{-5pt}\mbox{if }m-t \mbox{ is even}\\
\arg\min_{i\in\{1,2\}}(\pi_m(i)), & \hspace{-5pt}\mbox{if }m-t \mbox{ is odd}\\
\end{cases}, k_1,k_2=\emptyset\\
\end{cases}\nonumber\\
\end{eqnarray}
Thus the greedy decision is independent of feedback $f_{k_2}$ if $k_1<k_2$.
We now proceed to generalize equation (\ref{eq:ahat_2}). Let $k^*$ denote the latest slot for which an ARQ feedback is available from
\textit{one of the users} by slot $t$. Let $l=k^*-t-1$ for $k^*\neq \emptyset$ and $l=\emptyset$ for $k^*=\emptyset$ be a measure of
freshness of the latest ARQ feedback. Thus, using the preceding
discussion, we have
\begin{eqnarray}\label{eq:greedyk_2}
\lefteqn{\hat{a}_{t}|_{f_{k_1},f_{k_2},l_1,l_2,\vec{a}_{t+1},\pi_m}}\nonumber\\
&=&\begin{cases}
\begin{cases}
a_{k^*}, & \mbox{if }f_{k^*}=1\\
\bar{a}_{k^*}, & \mbox{if }f_{k^*}=0\\
\end{cases}, \mbox{if }k^*\neq\emptyset, k^*-t\mbox{ is even}\\
\begin{cases}
\bar{a}_{k^*}, & \mbox{if }f_{k^*}=1\\
a_{k^*}, & \mbox{if }f_{k^*}=0\\
\end{cases}, \mbox{if }k^*\neq\emptyset, k^*-t\mbox{ is odd}\\
\begin{cases}
\arg\max_{i\in\{1,2\}}(\pi_m(i)), & \mbox{if }m-t \mbox{ is even}\\
\arg\min_{i\in\{1,2\}}(\pi_m(i)), & \mbox{if }m-t \mbox{ is odd}\\
\end{cases},  k^*=\emptyset\\
\end{cases}\nonumber\\
\end{eqnarray}
where $\bar{a}_{k^*}$ is the user \textit{not} scheduled in slot $k^*$. This completes step $1$ of the
proof.

\textit{Step 2:}  
%We now proceed to show that,
If the greedy policy is implemented in slot $t$, 
%then the expected immediate reward in slot $t$ is
% independent of the scheduling decisions $\vec{a}_{t+1}$, i.e.,
% \begin{eqnarray}
% E_{\pi_{t}|\vec{a}_{t+1},\pi_m}R_{t}(\pi_t,\hat{a}_t)&=&E_{\pi_{t}|\pi_m}R_{t}(\pi_t,\hat{a}_t).
% \end{eqnarray}
the immediate reward expected in slot $t$, conditioned on scheduling decisions $\vec{a}_{t+1}$ and
initial belief $\pi_m$ can be rewritten as
\begin{eqnarray}\label{eq:exprewt_2}
\lefteqn{\E_{\pi_t|{\vec{a}_{t+1},\pi_m}}R_t(\pi_t,\hat{a}_t)}\nonumber\\
&=&\E_{\pi_t|{l=\emptyset,\vec{a}_{t+1},\pi_m}}(R_t(\pi_t,\hat{a}_t))P(l=\emptyset|{\vec{a}_{t+1},\pi_m})\nonumber\\
&&+\E_{l,l\neq\emptyset|{\vec{a}_{t+1},\pi_m}}\E_{\pi_t|{l,l\neq\emptyset,\vec{a}_{t+1},\pi_m}}(R_t(\pi_t,\hat{a}_t)).
\end{eqnarray}
Note that 
\begin{eqnarray}
\E_{\pi_t|{l=\emptyset,\vec{a}_{t+1},\pi_m}}(R_t(\pi_t,\hat{a}_t))&=&\max_{i}T^{(m-t)}(\pi_m(i))\nonumber\\
\end{eqnarray}
since, with $l=\emptyset$, i.e., no past feedback at the scheduler, the belief values at slot $t$ is independent of the past
scheduling decisions and is simply given by $\pi_t=T^{(m-t)}(\pi_m)$. Now rewriting the second part of
(\ref{eq:exprewt_2}),
\begin{eqnarray}\label{eq:EE_2}
\lefteqn{\E_{l,l\neq\emptyset|{\vec{a}_{t+1},\pi_m}}\E_{\pi_t|{l,l\neq\emptyset,\vec{a}_{t+1},\pi_m}}(R_t(\pi_t,\hat{a}_t))}\nonumber\\
&=&\E_{l,l\neq\emptyset|{\vec{a}_{t+1},\pi_m}}
\E_{\pi_{l+t+1}|{l,l\neq\emptyset,\vec{a}_{t+1},\pi_m}}\nonumber\\
&&\E_{\pi_t|{\pi_{l+t+1},l,l\neq\emptyset,\vec{a}_{t+1},\pi_m}}(R_t(\pi_t,\hat{a}_t)).
\end{eqnarray}
Consider $\E_{\pi_t|{\pi_{l+t+1},l,l\neq\emptyset,\vec{a}_{t+1},\pi_m}}(R_t(\pi_t,\hat{a}_t))$. From
the first step of the proof, the greedy decision in slot $t$ can be made solely based on the
latest feedback, i.e., $f_{k^*=l+t+1}$. This was recorded in (\ref{eq:greedyk_2}). 

Thus, when $l$ is an odd number (equivalently $k^*-t$ is even), if the feedback $f_{k^*}$ is an ACK (occurs with probability $\pi_{l+t+1}(a_{l+t+1})$) reschedule the user $a_{l+t+1}$ in slot
$t$. Conditioned on $f_{k^*}=1$, the belief value $\pi_{t}(a_{l+t+1})$ and hence the expected
immediate reward in slot $t$ is given by $T^l(p)$. If the feedback is a NACK, schedule the other
user denoted by $\bar{a}_{l+t+1}$. Conditioned on $f_{k^*}=0$, the belief value $\pi_{t}(\bar{a}_{l+t+1})$ and hence the 
expected immediate reward in slot $t$ is given by
$T^{(l+1)}(\pi_{l+t+1}(\bar{a}_{l+t+1}))=\pi_{l+t+1}(\bar{a}_{l+t+1})T^{l}(p)+(1-\pi_{l+t+1}(\bar{a}_{l+t+1}))T^{l}(r)$.
Averaging over $f_{k^*=l+t+1}$, when $l$ is odd,
\begin{eqnarray}
\lefteqn{\E_{\pi_t|{\pi_{l+t+1},l,l\neq\emptyset,\vec{a}_{t+1},\pi_m}}(R_t(\pi_t,\hat{a}_t))}\nonumber\\
&=&\pi_{l+t+1}(a_{l+t+1})T^{l}(p)+(1-\pi_{l+t+1}(a_{l+t+1}))\times\nonumber\\
&&\Big(\pi_{l+t+1}(\bar{a}_{l+t+1})T^{l}(p)+(1-\pi_{l+t+1}(\bar{a}_{l+t+1}))T^{l}(r)\Big)\nonumber\\
&=&P\big(\{S_{l+t+1}(1)=1 \cup S_{l+t+1}(2)=1\}|\nonumber\\
&&\hspace{96pt}{\pi_{l+t+1},l,l\neq\emptyset,\vec{a}_{t+1},\pi_m}\big)T^{l}(p)\nonumber\\
&&\hspace{-10pt}+P\big(\{S_{l+t+1}(1)=0 \cap S_{l+t+1}(2)=0\}|\nonumber\\
&&\hspace{96pt}{\pi_{l+t+1},l,l\neq\emptyset,\vec{a}_{t+1},\pi_m}\big)T^{l}(r)\nonumber\\
\end{eqnarray}
where $S_{k}(i)$ is the $1/0$ state of the channel of user $i$ in slot $k$. 

Using similar arguments, when $l$ is an even number,
\begin{eqnarray}
\lefteqn{\E_{\pi_t|{\pi_{l+t+1},l,l\neq\emptyset,\vec{a}_{t+1},\pi_m}}(R_t(\pi_t,\hat{a}_t))}\nonumber\\
&=&P\big(\{S_{l+t+1}(1)=0 \cup S_{l+t+1}(2)=0\}|\nonumber\\
&&\hspace{96pt}{\pi_{l+t+1},l,l\neq\emptyset,\vec{a}_{t+1},\pi_m}\big)T^{l}(r)\nonumber\\
&&\hspace{-10pt}+P\big(\{S_{l+t+1}(1)=1 \cap S_{l+t+1}(2)=1\}|\nonumber\\
&&\hspace{96pt}{\pi_{l+t+1},l,l\neq\emptyset,\vec{a}_{t+1},\pi_m}\big)T^{l}(p)\nonumber\\
\end{eqnarray}

Now, from (\ref{eq:EE_2}), using arguments similar to those used for (\ref{eq:rewfinal}) in the proof of Proposition~\ref{prop:myoopt}, we have
\begin{eqnarray}\label{eq:rewfinal_2}
\lefteqn{\E_{l,l\neq\emptyset|{\vec{a}_{t+1},\pi_m}}\E_{\pi_t|{l,l\neq\emptyset,\vec{a}_{t+1},\pi_m}}(R_t(\pi_t,\hat{a}_t))}\nonumber\\
&=&\E_{l,l\neq\emptyset|{\vec{a}_{t+1},\pi_m}}\nonumber\\
&&\begin{cases}
\hspace{-3pt}\Big(P\big(\{S_{l+t+1}(1)=1 \cup S_{l+t+1}(2)=1\}|{\pi_m}\big)T^{l}(p)\nonumber\\
\hspace{-3pt}+P\big(\{S_{l+t+1}(1)=0 \cap S_{l+t+1}(2)=0\}|{\pi_m}\big)T^{l}(r)\Big),\nonumber\\
\hspace{8pt}  \mbox{if } l \mbox{ is odd} \nonumber\\
\hspace{-3pt}\Big(P\big(\{S_{l+t+1}(1)=0 \cup S_{l+t+1}(2)=0\}|{\pi_m}\big)T^{l}(r)\nonumber\\
\hspace{-3pt}+P\big(\{S_{l+t+1}(1)=1 \cap S_{l+t+1}(2)=1\}|{\pi_m}\big)T^{l}(p)\Big), \nonumber\\
\hspace{8pt}  \mbox{if } l \mbox{ is even}
\end{cases}\nonumber\\
\end{eqnarray}
Now, along the lines of the proof of Proposition~\ref{prop:myoopt}, we average the expected reward over $l$ and obtain
\begin{eqnarray}
\lefteqn{\E_{\pi_t|{\vec{a}_{t+1},\pi_m}}R_t(\pi_t,\hat{a}_t)}\nonumber\\
&=&\hspace{-7pt}\prod_{k=m}^{t+1}P(D(a_k,k)>k-t-1)\max_{i}T^{(m-t)}(\pi_m(i))\nonumber\\
&&\hspace{-14pt}+\hspace{-7pt}\sum_{l=0}^{m-t-1}\hspace{-2pt}P(D(1,l+t+1)\le
l)\hspace{-7pt}\prod_{k=t+l}^{t+1}\hspace{-5pt}P(D(1,k)>k-t-1)\nonumber\\
&&\E_{l,l\neq\emptyset|{\vec{a}_{t+1},\pi_m}}\E_{\pi_t|{l,l\neq\emptyset,\vec{a}_{t+1},\pi_m}}(R_t(\pi_t,\hat{a}_t))
\end{eqnarray}
where the last quantity is given by (\ref{eq:rewfinal_2}). Thus the expected reward in slot $t$ is independent of the sequence of actions $\{a_{m},a_{m-1}\ldots
a_{t+1}\}$ if the greedy policy is implemented in slot $t$. By extension, the total reward expected from slot $t$ until the horizon is independent of the scheduling vector
$\vec{a}_{t+1}$ if the greedy policy is implemented in slots $\{t,t-1,\ldots, 1\}$, i.e.,  
\begin{eqnarray}
\sum_{k=t}^{1}\E_{\pi_k|{\vec{a}_{t+1},\pi_m}}R_k(\pi_k,\hat{a}_k)&=&\sum_{k=t}^{1}\E_{\pi_k|{\pi_m}}R_k(\pi_k,\hat{a}_k).\nonumber\\
\end{eqnarray}
Thus, if the greedy policy is optimal in slots $\{t,t-1,\ldots, 1\}$, then, it is also optimal in slot $t+1$.
Since $t$ is arbitrary and since the greedy policy is optimal at the horizon, by induction, the greedy
policy is optimal in every slot $\{m,m-1,\ldots, 1\}$. This establishes the proposition.

\section{Proof of Proposition~\ref{prop:counter}}\label{app:propcounter}

The proof proceeds in two steps: (1) We first construct a counterexample to the optimality of the greedy policy when the horizon, $m=4$ and arbitraty number of users $N; N>2$, (2) Based on this counterexample, we then construct a more general counterexample, with arbitraty $m; m>3$ and $N; N>2$. We proceed with the first step next. 

Assume an arbitrary number of users, $N; N>2$.  Let the horizon $m=4$.  Assume a deterministic ARQ delay of one time slot, i.e.,
$P_D(d=1)=1$ and $P_D(d\neq 1)=0$. Let the users be indexed in decreasing order of
their initial beliefs, i.e., $\pi_m(1)\ge \pi_m(2)\ge \ldots \pi_m(N)$. The net expected reward
corresponding to the greedy policy is given by
\begin{eqnarray}
\lefteqn{V_4(\pi_4,\{\hat{\textgoth{A}}_{k}\}_{k=1}^{4})}\nonumber\\
&=&\pi_4(1)+T(\pi_4(1))\nonumber\\
&&+\E_{f_4|\pi_4,a_4=1}[\hat{R}_{2}]+\E_{f_3,f_4|\pi_4,a_4=1,a_3=1}[\hat{R}_{1}]
\end{eqnarray}
Note that since the delay is one slot, the first ARQ feedback comes at the end of slot $3$.
Thus, the greedy decision in both slots $4$ and $3$ is user 1. Also, the greedy scheduler has access to
feedback $f_4$ only, at the beginning of slot $2$ and both feedback $f_4$ and $f_3$, at the beginning
of slot $1$. Therefore, $\hat{R}_{2}$ is averaged over $f_4$ and $\hat{R}_{1}$ is averaged over $f_4$
\textit{and} $f_3$. The average total reward under greedy policy can thus be evaluated by averaging
over all realizations of $f_4$ and $f_3$. Table~\ref{tab:greedy} lists the belief values of the three users 
in slots $2$ and $1$ for various values of $\{f_4,f_3\}$ along with the greedy decisions and
immediate rewards in slots $2$ and $1$. Note from the table that the belief value $\pi_2$ at
slot $2$ is a function of $f_4$ only, while $\pi_1$ at slot $1$ is a function of both $f_4$ and
$f_3$, consistent with the preceding discussion.

The probabilities of occurrence of the various realizations of $\{f_4,f_3\}$ are summarized below
\begin{eqnarray}
P(f_4,f_3)=\begin{cases}
\pi_4(1)p, & \mbox{if } \{f_4,f_3\}=\{1,1\}\\
\pi_4(1)(1-p), & \mbox{if } \{f_4,f_3\}=\{1,0\}\\
(1-\pi_4(1))r, & \mbox{if } \{f_4,f_3\}=\{0,1\}\\
(1-\pi_4(1))(1-r), & \mbox{if } \{f_4,f_3\}=\{0,0\}.
\end{cases}
\end{eqnarray}
Thus the net expected reward under the greedy policy is given by
\begin{eqnarray}
\lefteqn{V_4(\pi_4,\{\hat{\textgoth{A}}_{k}\}_{k=1}^{4})}\nonumber\\
&=&\hspace{-5pt}\pi_4(1)+T(\pi_4(1))+\pi_4(1)p\big(2T(p)\big)\nonumber\\
&&\hspace{-10pt}+\pi_4(1)(1-p)\big(T(p)+T^3(\pi_4(2))\big)\nonumber\\
&&\hspace{-10pt}+(1-\pi_4(1))r\big(T^2(\pi_4(2))+T(p)\big)\nonumber\\
&&\hspace{-10pt}+(1-\pi_4(1))(1-r)\big(T^2(\pi_4(2))+T^3(\pi_4(2))\big)
\end{eqnarray}

\begin{table*}
\begin{center}
\renewcommand{\tabcolsep}{.4cm}
\renewcommand{\arraystretch}{1.2}
\begin{tabular}{|c|c|c|c|c|c|c|}
\hline 
%\multicolumn{7}{|c|}{$p=0.9694,~r=0.1556 $} \\
% \multicolumn{4}{|c|}{$\pi_m=[0.1207~0.1962~0.1791]$} \\
% \multicolumn{4}{|c|}{$P_d=[0~1]$} \\
%\hline
& & & & & & \\
$\{f_4,f_3\}$ &  $\pi_2$ at slot $2$ & $\hat{a}_2$ & $\hat{R}_2$ & $\pi_1$ at
slot $1$ & $\hat{a}_1$ & $\hat{R}_1$ \\
& & & & & & \\
\hline
\{1,1\} & $\begin{bmatrix}T(p)\\T^2(\pi_4(2))\\T^2(\pi_4(3))\end{bmatrix}$ &  1 & $T(p)$ &
$\begin{bmatrix}T(p)\\T^3(\pi_4(2))\\T^3(\pi_4(3))\end{bmatrix}$ & 1 & $T(p)$\\
\hline
\{1,0\} & $\begin{bmatrix}T(p)\\T^2(\pi_4(2))\\T^2(\pi_4(3))\end{bmatrix}$ & 1 & $T(p)$ &
$\begin{bmatrix}T(r)\\T^3(\pi_4(2))\\T^3(\pi_4(3))\end{bmatrix}$ & 2 & $T^3(\pi_4(2))$\\
\hline
\{0,1\} & $\begin{bmatrix}T(r)\\T^2(\pi_4(2))\\T^2(\pi_4(3))\end{bmatrix}$ & 2 & $T^2(\pi_4(2))$ & 
$\begin{bmatrix}T(p)\\T^3(\pi_4(2))\\T^3(\pi_4(3))\end{bmatrix}$ & 1 & $T(p)$\\
\hline
\{0,0\} & $\begin{bmatrix}T(r)\\T^2(\pi_4(2))\\T^2(\pi_4(3))\end{bmatrix}$ & 2 & $T^2(\pi_4(2))$ &
$\begin{bmatrix}T(r)\\T^3(\pi_4(2))\\T^3(\pi_4(3))\end{bmatrix}$ & 2 & $T^3(\pi_4(2))$\\
\hline
\end{tabular}
\hspace{.1in}
\end{center}
\caption{Belief values, scheduling decisions, immediate rewards in slots $2$ and $1$ for various
realizations of ARQ feedback under the greedy policy.}
\label{tab:greedy}
\end{table*}

Now, with $a_k^*$ indicating the optimal decision in slot $k$, consider the following policy
$\tilde{\textgoth{A}}_k$ such that
$\tilde{a}_4=1,\tilde{a}_3=2,\tilde{a}_2=a_2^*,\tilde{a}_1=a_1^*$. Since the ARQ delay is
deterministic and equals one slot, the decision in slot $2$ does not affect the reward in slot $1$.
Thus the greedy policy is optimal in slot $2$. Trivially, greedy policy is optimal in slot $1$, as well.
Thus $a_2^*=\hat{a}_2$, $a_1^*=\hat{a}_1$. The average total reward under $\tilde{\textgoth{A}}_k$ is given by
\begin{eqnarray}
V_4(\pi_4,\{\tilde{\textgoth{A}}_{k}\}_{k=1}^{4})&=&\pi_4(1)+T(\pi_4(2))+\E_{f_4|\pi_4,a_4=1}[\tilde{R}_{2}]\nonumber\\
&&+\E_{f_3,f_4|\pi_4,a_4=1,a_3=2}[\tilde{R}_{1}]\nonumber\\
&=&\pi_4(1)+T(\pi_4(2))+\E_{f_4|\pi_4,a_4=1}[\hat{R}_{2}]\nonumber\\
&&+\E_{f_3,f_4|\pi_4,a_4=1,a_3=2}[\hat{R}_{1}]
\end{eqnarray}
We evaluate $V_4(\pi_4,\{\tilde{\textgoth{A}}_{k}\}_{k=1}^{4})$ along the lines of the greedy
net expected reward evaluation. Table~\ref{tab:arbi} summarizes the beliefs, scheduling decision $\tilde{a}_k$ and immediate
rewards in slots $2$ and $1$ for all the realizations of $\{f_4,f_3\}$ when
$\{\tilde{a}_4,\tilde{a}_3\}=\{1,2\}$. Users are once again ordered according to their initial
belief values, i.e., $\pi_4(1)\ge\pi_4(2)\ge \pi_4(3)$. Note from the table that the belief value $\pi_2$ at
slot $2$ is a function of $f_4$ only, while $\pi_1$ at slot $1$ is a function of both $f_4$ and
$f_3$, consistent with the ARQ delay profile.

\begin{table*}
\begin{center}
\renewcommand{\tabcolsep}{.4cm}
\renewcommand{\arraystretch}{1.2}
\begin{tabular}{|c|c|c|c|c|c|c|}
\hline 
%\multicolumn{7}{|c|}{$p=0.9694,~r=0.1556 $} \\
% \multicolumn{4}{|c|}{$\pi_m=[0.1207~0.1962~0.1791]$} \\
% \multicolumn{4}{|c|}{$P_d=[0~1]$} \\
%\hline
& & & & & & \\
$\{f_4,f_3\}$ &  $\pi_2$ at slot $2$ & $\tilde{a}_2$ & $\tilde{R}_2$ & $\pi_1$ at
slot $1$ & $\tilde{a}_1$ & $\tilde{R}_1$ \\
& & & & & & \\
\hline
\{1,1\} & $\begin{bmatrix}T(p)\\T^2(\pi_4(2))\\T^2(\pi_4(3))\end{bmatrix}$ &  1 & $T(p)$ &
$\begin{bmatrix}T^2(p)\\T(p)\\T^3(\pi_4(3))\end{bmatrix}$ & 2 & $T(p)$\\
\hline
\{1,0\} & $\begin{bmatrix}T(p)\\T^2(\pi_4(2))\\T^2(\pi_4(3))\end{bmatrix}$ & 1 & $T(p)$ &
$\begin{bmatrix}T^2(p)\\T(r)\\T^3(\pi_4(3))\end{bmatrix}$ & 1 & $T^2(p)$\\
\hline
\{0,1\} & $\begin{bmatrix}T(r)\\T^2(\pi_4(2))\\T^2(\pi_4(3))\end{bmatrix}$ & 2 & $T^2(\pi_4(2))$ & 
$\begin{bmatrix}T^2(r)\\T(p)\\T^3(\pi_4(3))\end{bmatrix}$ & 2 & $T(p)$\\
\hline
\{0,0\} & $\begin{bmatrix}T(r)\\T^2(\pi_4(2))\\T^2(\pi_4(3))\end{bmatrix}$ & 2 & $T^2(\pi_4(2))$ &
$\begin{bmatrix}T^2(r)\\T(r)\\T^3(\pi_4(3))\end{bmatrix}$ & 3 & $T^3(\pi_4(3))$\\
\hline
\end{tabular}
\hspace{.1in}
\end{center}
\caption{Belief values, scheduling decisions, immediate rewards in slots $2$ and $1$ for various
realizations of ARQ feedback under policy $\tilde{\textgoth{A}}_{k}$.}
\label{tab:arbi}
\end{table*}

\begin{table*}
\begin{center}
\renewcommand{\tabcolsep}{.32cm}
\renewcommand{\arraystretch}{1.25}
\begin{tabular}{|c|c|c|c|c|c|}
\hline 
%\multicolumn{7}{|c|}{$p=0.9694,~r=0.1556 $} \\
% \multicolumn{4}{|c|}{$\pi_m=[0.1207~0.1962~0.1791]$} \\
% \multicolumn{4}{|c|}{$P_d=[0~1]$} \\
%\hline
& & & & & \\
$p$ & $r$ & $\pi_4$ & $V_4(\pi_4,\{\tilde{\textgoth{A}}\}_{k=1}^{4})$ & $V_4(\pi_4,\{\hat{\textgoth{A}}\}_{k=1}^{4})$ & $V_4(\pi_4,\{\tilde{\textgoth{A}}\}_{k=1}^{4})$\\
& & & & & $-V_4(\pi_4,\{\hat{\textgoth{A}}\}_{k=1}^{4})$\\
\hline
0.9308 & 0.1797 & $\begin{bmatrix}0.5216\\0.5130\\0.3305\end{bmatrix}$ & 2.6368 & 2.6141  & 0.0227\\
\hline
0.8875 & 0.0186 & $\begin{bmatrix}0.3416\\0.3310\\0.2648\end{bmatrix}$ & 1.6155 & 1.5454  & 0.0701\\
\hline
\end{tabular}
\hspace{.1in}
\end{center}
\caption{Sample system parameters when the greedy policy is suboptimal. Number of users $N=3$, deterministic delay $D=1$, horizon
$m=4$ is used.}
\label{tab:param}
\end{table*}

The probabilities of occurrence of the various realizations of $\{f_4,f_3\}$ when $a_4=1,a_3=2$,
are summarized below.
\begin{eqnarray}
\lefteqn{P(f_4,f_3)}\nonumber\\
&=&\hspace{-8pt}\begin{cases}
\pi_4(1)T(\pi_4(2)), & \mbox{if } \{f_4,f_3\}=\{1,1\}\\
\pi_4(1)(1-T(\pi_4(2))), & \mbox{if } \{f_4,f_3\}=\{1,0\}\\
(1-\pi_4(1))T(\pi_4(2)), & \mbox{if } \{f_4,f_3\}=\{0,1\}\\
(1-\pi_4(1))(1-T(\pi_4(2))), & \mbox{if } \{f_4,f_3\}=\{0,0\}.
\end{cases}\nonumber\\
\end{eqnarray}
Thus, the net expected reward under policy $\tilde{\textgoth{A}}_k$ is given by
\begin{eqnarray}\label{eq:diff0}
\lefteqn{V_4(\pi_4,\{\tilde{\textgoth{A}}\}_{k=1}^{4})}\nonumber\\
&=&\pi_4(1)+T(\pi_4(2))+\pi_4(1)T(\pi_4(2))\big(2T(p)\big)\nonumber\\
&&+\pi_4(1)(1-T(\pi_4(2)))\big(T(p)+T^2(p)\big)\nonumber\\
&&+(1-\pi_4(1))T(\pi_4(2))\big(T^2(\pi_4(2))+T(p)\big)\nonumber\\
&&+(1-\pi_4(1))(1-T(\pi_4(2)))\big(T^2(\pi_4(2))+T^3(\pi_4(3))\big)\nonumber\\
\end{eqnarray}
We now proceed to show that, for $N=3$, deterministic ARQ delay $D=1$ and horizon $m=4$,
$\exists~p,r,\pi_4$ such that the net expected reward corresponding to policy
$\tilde{\textgoth{A}}_k$ is strictly higher than that of the greedy policy. The difference in
reward, after algebraic manipulations is given by
\begin{eqnarray}\label{eq:diff1}
\lefteqn{V_4(\pi_4,\{\tilde{\textgoth{A}}\}_{k=1}^{4})-V_4(\pi_4,\{\hat{\textgoth{A}}\}_{k=1}^{4})}\nonumber\\
&=&(p-r)\Big(\pi_4(2)-\pi_4(1)\nonumber\\
&&+(p-r)^2(1-\pi_4(1))\pi_4(3)\big(1-r-(p-r)\pi_4(2)\big)\Big).\nonumber\\
\end{eqnarray}
For the special case $\pi_4(1)=\pi_4(2)=\frac{1}{2}$, we have
\begin{eqnarray}
\lefteqn{V_4(\pi_4,\{\tilde{\textgoth{A}}\}_{k=1}^{4})-V_4(\pi_4,\{\hat{\textgoth{A}}\}_{k=1}^{4})}\nonumber\\
&=&\frac{(p-r)^3}{2}\Big(1-\frac{p+r}{2}\Big)\pi_4(3)
\end{eqnarray}
For any $p<1$, since $p>r$,
$V_4(\pi_4,\{\tilde{\textgoth{A}}\}_{k=1}^{4})>V_4(\pi_4,\{\hat{\textgoth{A}}\}_{k=1}^{4})$
$\forall~\pi_4(3)>0$. With the net expected reward of the optimal policy being no less than
$V_4(\pi_4,\{\tilde{\textgoth{A}}\}_{k=1}^{4})$, we see that the greedy policy is not in general
optimal. Table~\ref{tab:param} lists a few other values of $p,r,\pi_4$ for which the greedy policy is suboptimal. This
establishes a counterexample for the optimality of the greedy policy when $N>2$. 

A more general counterexample for an arbitrary horizon length $m$ can be constructed based on the one thus established. We proceed with this constructuion in the sequel.
As before, asssume $\pi_m(1)\ge\pi_m(2)\ge \ldots \pi_m(N)$ and  a deterministic ARQ delay of one time slot, i.e.,
$P_D(d=1)=1$. With $S_k(i)$ indicating the underlying state of the channel of user $i$ in slot $k$, consider the following realization of the channel: 
$\mathcal{R}_1 =\{S_m(1)=1,S_{m-1}(1)=1,\ldots S_5(1)=1; S_m(2)=1,S_{m-1}(2)=1,\ldots S_{5}(2)=1\}$. Recall the policy $\tilde{\textgoth{A}}$ from above. We define a variant of this policy, $\tilde{\textgoth{B}}$, as follows. Policy $\tilde{\textgoth{B}}$ performs greedy scheduling in slots $\{m\ldots 5\}$. Under realization $\mathcal{R}_1$, policy $\tilde{\textgoth{B}}$, being a greedy scheduler, schedules user 1 in slots $\{m\ldots 5\}$. Thus the realization $\{S_m(1)=1,S_{m-1}(1)=1,\ldots S_6(1)=1\}$ is observable by policy $\tilde{\textgoth{B}}$ by the beginning of slot $4$. From slot $4$, under realization $\mathcal{R}_1$, define policy $\tilde{\textgoth{B}}$ such that it behaves along the lines of policy $\tilde{\textgoth{A}}$ defined earlier. Also, policy  $\tilde{\textgoth{B}}$ performs greedy scheduling in all slots $\{m\ldots 1\}$ under all channel realizations other than $\mathcal{R}_1$. Thus the reward difference between policy $\tilde{\textgoth{B}}$ and the greedy scheduler $\hat{\textgoth{A}}$ is given by the difference in slots $\{4\ldots 1\}$, under realization $\mathcal{R}_1$, weighted by the probability of this realization. Formally,
\begin{eqnarray}\label{eq:diff2}
\lefteqn{V_m(\pi_m,\{\tilde{\textgoth{B}}\}_{k=1}^{m})-V_m(\pi_m,\{\hat{\textgoth{A}}\}_{k=1}^{m})}\nonumber\\
&=&\textrm{Prob}\{\mathcal{R}_1\}\times\nonumber\\
&&\big(V_4(\pi_4,\{\tilde{\textgoth{A}}\}_{k=1}^{4})-V_4(\pi_4,\{\hat{\textgoth{A}}\}_{k=1}^{4})\big)
\end{eqnarray}
with $\pi_4=\{p,p,T^{m-4}(\pi_m(3)),\ldots T^{m-4}(\pi_m(N))\}$. Note that the belief values $\pi_4$ reflect the realization $\mathcal{R}_1$ and the greedy nature of both policies until slot $5$. Now, since policy $\tilde{\textgoth{B}}$ is defined to behave like policy $\tilde{\textgoth{A}}$ from slot $4$, we can use the reward difference expression in (\ref{eq:diff1}) to simplify (\ref{eq:diff2}), as below.
\begin{eqnarray}
\lefteqn{V_m(\pi_m,\{\tilde{\textgoth{B}}\}_{k=1}^{m})-V_m(\pi_m,\{\hat{\textgoth{A}}\}_{k=1}^{m})}\nonumber\\
&=&\textrm{Prob}\{\mathcal{R}_1\}\Big((p-r)\Big(\pi_4(2)-\pi_4(1)\nonumber\\
&&+(p-r)^2(1-\pi_4(1))\pi_4(3)\times\nonumber\\
&&\big(1-r-(p-r)\pi_4(2)\big)\Big)\Big)|_{\pi_4(1)=\pi_4(2)=p}\nonumber\\
&=&\textrm{Prob}\{\mathcal{R}_1\}\Big((p-r)^3(1-p)((1-r)-(p-r)p)\Big)\nonumber\\
\end{eqnarray}
Note that the last equality is always positive rendering the greedy policy $\hat{\textgoth{A}}$ suboptimal. This establishes a more general counterexample to the optimality of the greedy policy when $N>2$. The proposition is thus proved.
 
\section{Proof of Proposition~\ref{prop:4pt5}}\label{app:prop4pt5}

\begin{table*}
\begin{center}
\renewcommand{\tabcolsep}{.32cm}
\renewcommand{\arraystretch}{1.25}
\begin{tabular}{|c|c|c|c|c|c|c|c|}
\hline 
%\multicolumn{7}{|c|}{$p=0.9694,~r=0.1556 $} \\
% \multicolumn{4}{|c|}{$\pi_m=[0.1207~0.1962~0.1791]$} \\
% \multicolumn{4}{|c|}{$P_d=[0~1]$} \\
%\hline
%& & & & & \\
$p_1=p_2$ & $r_1$ & $r_2$ & $\pi_1$ & $\pi_2$ & $T(\pi_2(1))$ & $T(\pi_2(2))$ &  $\hat{V}_2-\tilde{V}_2$\\
\hline
0.5060 & 0.1411 & 0.1054 & 0.2276 & 0.2179 & 0.2241 & 0.1926 & -0.0119\\
\hline
0.6333 & 0.3952 & 0.1296 & 0.5864 & 0.5861 & 0.5348 & 0.4248 & -0.0452\\
\hline
\end{tabular}
\hspace{.1in}
\end{center}
\caption{Sample system parameters when the greedy policy is suboptimal under non-identical Markov channels. Number of users $N=2$, instantaneous ARQ feedback $D=0$, horizon
$m=2$ are assumed.}
\label{tab:nonid}
\end{table*}

Consider the case when $N=2$ and the ARQ feedback is instantaneous (end of slot), i.e., $D=0$. Let $p_i,r_i$ indicate the Markov channel probabilities for user $i\in\{1,2\}$. Let $p_1=p_2>r_1>r_2$. Assume horizon length $m=2$. Let $\pi_2(1)>\pi_2(2)$. The total reward under greedy decision in the current slot $k=2$ and optimal reward at the horizon, i.e., $k=1$, is given by 
\begin{eqnarray}
\hat{V}_2&=&\pi_2(1)+\pi_2(1)\max(p_1,T(\pi_2(2)))\nonumber\\
&&+(1-\pi_2(1))\max(r_1,T(\pi_2(2)))\nonumber\\
&=&\pi_2(1)+\pi_2(1)p_1+(1-\pi_2(1))\max(r_1,T(\pi_2(2)))\nonumber\\
\end{eqnarray}
where we have used the fact that the greedy policy is optimal in the last slot and $p_1>T(\pi_2(1))>T(\pi_2(2))$.

The total reward when the non-greedy decision is made in the current slot and optimal decision is made in the last slot is given by
\begin{eqnarray}
\tilde{V}_2&=&\pi_2(2)+\pi_2(2)\max(p_2,T(\pi_2(1)))\nonumber\\
&&+(1-\pi_2(2))\max(r_2,T(\pi_2(1)))\nonumber\\
&=&\pi_2(2)+\pi_2(2)p_2+(1-\pi_2(2))\max(r_1,T(\pi_2(1)))\nonumber\\
\end{eqnarray}
where we have used $p_2=p_1>T(\pi_2(1))$ and $r_2<T(\pi_2(2))<T(\pi_2(1))$.
Now considering the special case when $T(\pi_2(2))>r_1$, we have, with algebraic manipulations,
\begin{eqnarray}
\hat{V}_2-\tilde{V}_2&=&(\pi_2(1)-\pi_2(2))\nonumber\\
&&-(r_1-r_2)(1-\pi_2(1))(1-\pi_2(2)).
\end{eqnarray}
In Table~\ref{tab:nonid}, we provide numerical examples consistent with the setup assumed above, yielding negative values for $\hat{V}_2-\tilde{V}_2$. This establishes that the greedy policy is not, in general, optimal, when the Markov channels are non-identical, even when the number of users, $N=2$ and the ARQ delay is instantaneous. The proposition is thus proved.

\section{Key Quantities}\label{app:key}
\begin{tabular}{p{15pt}p{3pt}p{200pt} }

N &: & Number of users in the downlink\\
$p$ &: & $P(\textrm{channel is ON in the current slot}~|$\\
&& \hspace{35pt}$\textrm{channel was ON in the previous slot})$\\
$r$ &: & $P(\textrm{channel is ON in the current slot}~|$\\
&& \hspace{35pt}$\textrm{channel was OFF in the previous slot})$\\
$m$ &: & Horizon\\
$\pi_{t}(i)$ &: & Belief value of user $i$ in slot $t$\\
$T^u(.)$ &: & $u$-step belief evolution operator\\
$a_k$ &: & Index of the user scheduled in slot $k$\\
$\textgoth{A}_{k}$ &: & Scheduling policy applied in slot $k$\\
$\widehat{\textgoth{A}}_{k}$ &: & Greedy scheduling policy applied in slot $k$\\
$f_k$ &: & Feedback originating from slot $k$\\
$F_t$ &: & Feedback arriving at slot $t$\\
$D(i,t)$ &: & Delay of feedback from user $i$ in slot $t$\\
$P_D(.)$ &: & Probability mass function of \emph{i.i.d} delay $D$\\
$V_t$ &: & Net expected reward in slot $t$\\
%$\eta_{\summ}(m,\{\textgoth{A}_k\}^m_{k=1})$ &: & Sum throughput of the downlink\\
$C_{\summ}$ &: & Sum capacity of the downlink\\
$C_{\summ}^{\genie}$ &: & Sum capacity of the genie-aided downlink\\
$\mu_i^{\textgoth{A}}$ &: & Throughput of user $i$ under scheduling policy $\textgoth{A}$
%
%
% $R_{k}$ &: & Expected current reward in the control interval $k$\\
% $V_k$ &: & Net expected reward in the control interval $k$\\
% $\eta_{\textrm{sum}}$ &: & Sum throughput\\
% $\textgoth{A}_{k}$ &: & Scheduling policy applied in the control interval $k$\\
% $\widehat{\textgoth{A}}_{k}$ &: & Greedy scheduling policy applied in the control interval $k$\\
% $\textgoth{A}^{*}_{k}$ &: & Optimal scheduling policy\\
% $S_{k}$ &: & State vector such that $S_{k}(i)$ indicates the state (1-ON/0-OFF) of the channel of user
% $i$ in control interval $k$\\
% $O_{k}$ &: & Schedule order vector in the control interval $k$, the ordered arrangement of the index of the users in decreasing
% order of $\pi_{k}(i)$\\
% $C_{\summ}$ &: & Sum capacity of the downlink environment\\
% $p_s$ &: & Steady state ON probability of the Markov channels\\
%
\end{tabular}

\begin{biography}{Sugumar Murugesan} (S10 / M11)
received the B.E. degree in Electronics and Communication Engineering from
the College of Engineering, Anna University, India in 2004 and the M.S. and Ph.D. degrees in Electrical
and Computer Engineering from the Ohio State University, USA, in 2006 and 2010, respectively. He is a recipient of the OSU University Fellowship (2004-05). Dr. Murugesan is currently a Post-doctoral Research Associate with the Department of ECEE at Arizona State University. His research interests include communication theory, wireless networks, sequential decision processes and smart power grids.
\end{biography}
\vspace{-10pt}

\begin{biography}{Philip Schniter} (S03 / M93 / SM05) received the
B.S. and M.S. degrees in electrical and computer
engineering from the University of Illinois at Urbana-
Champaign in 1992 and 1993, respectively.
In 2000, he received the Ph.D. degree in electrical
engineering from Cornell University, Ithaca, NY.
From 1993 to 1996, he was with Tektronix Inc.,
Beaverton, OR, as a Systems Engineer. Subsequently,
he joined the Department of Electrical and
Computer Engineering, The Ohio State University,
Columbus, OH, where he is now an Associate
Professor. His research interests include signal processing, communication
theory, and wireless networks.
Dr. Schniter received the National Science Foundation CAREER Award in
2003.
\end{biography}
\vspace{-10pt}

\begin{biography}{Ness B. Shroff} (S91 / M93 / SM01/ F07) currently holds 
 the Ohio Eminent Scholar Chaired Professorship in Networking
and Communications, in the departments of ECE and
CSE at The Ohio State University. He is also a Guest Chaired Professor of Wireless Networking in the department of Electronic Engineering  at Tsinghua University, China.  Previously, he
was a Professor of ECE at Purdue University and
the director of the Center for Wireless Systems
and Applications (CWSA), a university-wide center
on wireless systems and applications. His research
interests span the areas of wireless and wireline communication
networks, where he investigates fundamental
problems in the design, performance, pricing,
and security of these networks. Dr. Shroff has received numerous awards
for his networking research, including the NSF CAREER award, the best
paper awards for IEEE INFOCOM 06 and IEEE INFOCOM08, the best paper
award for IEEE IWQoS06, the best paper of the year award for the Computer
Networks journal, and the best paper of the year award for the Journal of
Communications and Networks (JCN) (his IEEE INFOCOM05 paper was
one of two runner-up papers).
\end{biography}

\end{document}